\documentclass[utf8]{FrontiersinHarvard}

\usepackage{url,hyperref,microtype,subcaption}
\usepackage[onehalfspacing]{setspace}
\usepackage{amsmath,amsfonts}
\usepackage{algorithmic}
\usepackage{array}
\usepackage[caption=false,font=normalsize,labelfont=sf,textfont=sf]{subfig}

\makeatletter
\newenvironment{description}
  {\list{}{\labelwidth\z@ \itemindent-\leftmargin
   }}
  {\endlist}

\makeatother
\usepackage{enumitem}

\usepackage{textcomp}
\usepackage{stfloats}
\usepackage{verbatim}
\usepackage{graphicx}
\def\BibTeX{{\rm B\kern-.05em{\sc i\kern-.025em b}\kern-.08em
    T\kern-.1667em\lower.7ex\hbox{E}\kern-.125emX}}
\usepackage{balance}
\usepackage{multirow}
\usepackage{booktabs}
\usepackage{makecell}
\usepackage{amssymb}
\usepackage{pdfpages}
\newcommand{\etal}{et al. }

\def\keyFont{\fontsize{8}{11}\helveticabold }
\def\firstAuthorLast{Hu {et~al.}} 
\def\Authors{Yong-Hao Hu\,$^{1,2}$, Sotaro Yokoi\,$^{1,2}$, Yuji Hatada\,$^{3}$, Yuichi Hiroi\,$^{2}$, Takuji Narumi\,$^{1}$ and Takefumi Hiraki\,$^{2,4,*}$}
\def\Address{$^{1}$Graduate School of Information Science and Technology, The University of Tokyo, Tokyo, Japan \\
$^{2}$Cluster Metaverse Lab, Tokyo, Japan \\
$^{3}$Interfaculty Initiative in Information Studies, The University of Tokyo, Tokyo, Japan \\
$^{4}$Institute of Library, Information and Media Science, University of Tsukuba, Ibaraki, Japan
}
\def\corrAuthor{Yong-Hao Hu}
\def\corrEmail{yh-haoareyou@cyber.t.u-tokyo.ac.jp}

\begin{document}
\onecolumn
\firstpage{1}

\title[LUIDA: Experimentation Framework on Metaverse]{LUIDA: Large-scale Unified Infrastructure for Digital Assessments based on\\Commercial Metaverse Platform} 

\author[\firstAuthorLast ]{\Authors} 
\address{} 
\correspondance{} 

\extraAuth{}

\maketitle

\begin{abstract}
Online experiments using metaverse platforms have gained significant traction in Human-Computer Interaction and Virtual Reality (VR) research. However, current research workflows are highly fragmented, as researchers must use separate tools for system implementation, participant recruitment, experiment execution, and data collection, reducing consistency and increasing workload. Meanwhile, online experimental approaches based on metaverse platforms have been proposed and explored. However, the lack of a structured methodology forces researchers to either acquire specialized expertise needed for each different platform, or resort to inefficient, manual methods for their experiments. We present LUIDA (Large-scale Unified Infrastructure for Digital Assessments), a metaverse-based framework that integrates these fragmented processes. LUIDA automatically allocates interconnected virtual environments for parallel experiment execution and provides implementation templates adaptable to various VR research domains, requiring minimal metaverse development expertise. Our evaluation included two studies using a prototype built on Cluster, the commercial metaverse platform. First, VR researchers using LUIDA to develop and run experiments reported high usability scores (SUS: 73.75) and moderate workload (NASA-TLX: 24.11) for overall usage, with interviews confirming streamlined workflows compared to traditional laboratory experiments. Second, we conducted three replicated experiments with public Cluster users, each recruiting approximately 200 participants within one week. These experiments produced results that closely matched the original studies, validating the experimental integrity of LUIDA across research domains. LUIDA has been released publicly and is undergoing technical refinements, eventually providing a unified and structured protocol to improve research efficiency and experimental reproducibility in VR studies.

\tiny
 \keyFont{ \section{Keywords:} Metaverse, Social VR, Remote experiment, Experiment platform, User study} 

\end{abstract}

\section{Introduction}

The advancement of experimental research in Human-Computer Interaction (HCI) and Virtual Reality (VR) has been accompanied by various tools and platforms, which remain highly fragmented:
many of them target specific research domains~\cite{rocketbox, hart} or isolated procedures such as participant recruitment~\cite{jikken_baito, sona} or data collection~\cite{bmlTUX}.
Such fragmentation forces researchers to juggle multiple systems throughout the experimental workflow, compromising experimental design consistency and increasing workflow complexity.

Simultaneously, due to the growth of crowd-sourcing platforms such as Amazon Mechanical Turk\footnote{\url{https://www.mturk.com/}}, online experimental systems are gaining significant attention as alternatives or supplements to traditional in-person experiments, particularly since the outbreak of the COVID-19 pandemic in 2020.
In the field of VR research, the possibility of online experimentation has also been demonstrated by the growing use of metaverse platforms such as VRChat\footnote{\url{https://hello.vrchat.com/}}.
Studies have demonstrated that experimental configurations originally designed for physical settings can be effectively transferred to virtual environments~\cite{remote_vr_exp_framework, remote_exp_social_vr}, enabling access to a broader, more diverse participant pool.
However, despite their promise, these online approaches come with significant limitations.
Technical barriers persist, as crowd-sourcing platforms require expertise in networking or experiment automation;
current metaverse attempts are restricted by limited customizability or data collection methods.
Furthermore, the lack of a structured methodology burdens researchers with acquiring platform-specific expertise or forcing them to fall back on inefficient, manual methods.
For instance, researchers implementing their experimental system with available toolkits (e.g., bmlTUX~\cite{bmlTUX}) are still required to find participant pools via recruitment platforms (e.g., Sona~\cite{sona}) or implement additional programs to retrieve data over the internet;
experiments in social VR necessitate that researchers manually record data instead of using automatic logging~\cite{remote_exp_social_vr}.

To address these challenges, we propose the experimental framework LUIDA (Large-scale Unified Infrastructure for Digital Assessments).
Its novelty lies in (1) integrating typically fragmented experimental workflows, from recruitment to data collection, and (2) establishing a structured methodology for metaverse-based experimentation by leveraging inherent platform capabilities for automation and parallelization, while abstracting technical complexities via an adaptable low-code template.
Specifically, LUIDA conducts participant recruitment directly within the metaverse via an in-world bulletin board;
automates and parallelizes experiment execution by leveraging the metaverse's capacity to host multiple virtual spaces for simultaneous sessions;
provides automated, continuous motion tracking with customizable data collection uploaded to online storage;
and offers a low-code Unity-based template for experiment implementation, with graphical interfaces for variables and state machine configurations.

LUIDA is designed with a platform-agnostic philosophy, allowing adaptation to different metaverse platforms.
In this paper, we develop a prototype using the commercial metaverse platform Cluster to leverage its existing user base and content creation tools.
To evaluate LUIDA's effectiveness, we performed a two-step validation: a user study with VR researchers replicating representative experiments from perception, cognition, and user interface domains, followed by public deployment on Cluster with general participants.

Our main contributions include the following:
\begin{description}[leftmargin=0pt,labelindent=10pt,itemsep=0pt]
    \item[Unified experimental framework:] This study proposed LUIDA, a framework that unifies fragmented VR experimental workflows by providing efficient solutions, including system implementation across multiple research domains, and automated, parallel participant recruitment and data collection.
    \item[Structured methodology for metaverse-based research:] LUIDA provides a novel, structured methodology that requires minimal platform-specific expertise and removes reliance on inefficient, manual approaches.
    \item[Validation of experimental validity:] We conducted user studies with VR researchers to use LUIDA, as well as online experiments with nearly 200 participants per experiment, a total of over 500 datasets in one week.
    \item[Pathway to an open platform:] We identified limitations and future development pathways of LUIDA to release it as an open platform.
\end{description}

The remainder of this paper is structured as follows.
Section~\ref{sec:related-work} supports LUIDA's novelty by reviewing the relevant literature, comparing LUIDA with current tools and prior metaverse-based attempts in addressing desirable functionalities for VR researchers.
Sections~\ref{sec:overview} and~\ref{sec:core-features} detail the system architecture and core mechanisms designed to address research needs identified in Section~\ref{sec:related-work}.
Section~\ref{sec:exp-to-replicate} outlines the evaluation design using three representative experiments,
while Sections~\ref{sec:user-study-with-researchers} and~\ref{sec:user-study-online} report on a usability study with VR researchers and the subsequent online replication experiments.
Finally, Section~\ref{sec:discussion} discusses the findings, and Section~\ref{sec:conclusion} concludes the research.

\section{Related Work}
\label{sec:related-work}
\subsection{Desirable Functionalities for Comprehensive VR Research}
\label{subsec:related:desirable-func}
Based on a review of previous literature and existing tools, we identified desirable functionalities to improve efficiency, quality and validity of VR studies.

\subsubsection{Participant Recruitment}

\paragraph{Larger and More Diverse User Pools}
\label{par:desirable-func-larger_and_more_diverse_user_pools}
VR studies, especially those of psychology or cognition, increasingly require large sample sizes of dozens or hundreds of participants~\cite{proteus_review, Makransky2021TheCA, MORENO2019834}, yet traditional laboratory methods face time and resource constraints. Reviews have noted that small, homogeneous samples result in underpowered studies with limited generalizability~\cite{vr_exp_check_statistical_power, vr_exp_validity_challenges}.
Larger, diverse pools are desirable to enhance statistical power and external validity~\cite{vr_exp_check_statistical_power, vr_exp_validity_challenges}.

\paragraph{Access to VR-experienced Users}
\label{par:desirable-func-vr-users}
Traditional crowdsourcing tools struggle to recruit sufficient VR-experienced users~\cite{crowdsourcing_vr_experiments}, but metaverse platforms provide access to a global pool of users who already own the necessary hardware~\cite{recruit_on_vrchat}.
These platforms also enable ``in-the-wild'' studies observing naturalistic social behaviors, effectively serving as a ``living laboratory'' for complex human interactions impossible to replicate in controlled settings~\cite{vr_ethnography, vr_as_research_tool}.

\subsubsection{Experiment Execution}

\paragraph{Multi-Stage Experiment Management}
\label{par:desirable-func-multi-stage}
VR experiments often involve multiple stages, including avatar familiarization~\cite{vrpt_van_Loon_Bailenson, Waltemate2018_avatar, VEQ}, in-VR questionnaires~\cite{putze_questionnaire}, and tasks before or after the main trials~\cite{neurorehabilitation}.
They may also include consecutive scenarios like training~\cite{Gavish2015_industrial_training} or perspective-taking tasks~\cite{vrpt_van_Loon_Bailenson}.
However, a unified protocol for managing such stages is currently lacking.

\paragraph{Avatar Control and Investigation}
\label{par:desirable-func-avatar}
Avatars are users' virtual representations in immersive environments.
They foster a sense of embodiment that influences users' emotions, behavior, and even self-perception~\cite{embodiment}, making them powerful experimental variables.
Research is expanding on how they affect perception~\cite{virtual_hand_affordance} and the Proteus effect, where avatar appearance influences users' attitudes or behavior~\cite{proteus, proteus_review}.

\paragraph{Asynchronized Execution for Experiments in Larger Scales}
\label{par:desirable-func-async-exec}
Traditional 1-on-1 lab sessions limit efficiency and participant recruitment.
While remote, asynchronous approaches exist through distributing software or hardware, they require significant resources, logistical management, and technical expertise on networking~\cite{remote_vr_exp_framework, remote_vr_exp_Steed}.
Metaverse platforms offer a solution by providing access to users who already own VR devices, enabling asynchronous, parallel execution that improves both efficiency and external validity.

\subsubsection{Data Collection}

\paragraph{Immersive Questionnaire Administration}
\label{par:desirable-func-questionnaire}

Administering questionnaires within VR enhances the consistency of the sense of presence~\cite{schwind_presence_questionnaire} and prevents immersion breaks caused by switching between virtual and real space~\cite{putze_questionnaire}.
These effects benefit studies on presence~\cite{IPQ, schwind_presence_questionnaire}, embodiment~\cite{VEQ, Genay2022}, applications~\cite{vr_rehabilitation}, and so on.

\paragraph{Real-time Behavioral Tracking}
\label{par:desirable-func-behavioral-tracking}
Continuous capture of users' positions and postures enables quantification of subtle behavioral patterns, such as proxemics~\cite{Yaremych2019TracingPB} and movement variations influenced by avatar use~\cite{drumming_proteus}.

\paragraph{Integration with External Hardware}
\label{par:desirable-func-hardware}

External hardware enables the collection of biometric data~\cite{biofeedback}, EMG signals~\cite{emg}, and eye-gaze data~\cite{eye_tracking}, which are difficult to capture using VR headsets alone.
These devices can also provide outputs such as haptic feedback~\cite{haptics} or full-body tracking for enhanced embodiment or behavioral tracking~\cite{motion_tracker}.

\paragraph{Data Integrity and Privacy}
\label{par:desirable-func-data-integrity}

Unsupervised remote experiments face issues like participants' non-compliance and low-quality data that threaten data integrity~\cite{vr_exp_validity_challenges, in_the_wild_exp}, which may be mitigated through device verification or rigorous data filtering~\cite{crowdsourcing_vr_experiments, in_the_wild_exp}.
Furthermore, privacy concerns arise from collecting sensitive data (e.g., private information, biometric data) and transmitting it online~\cite{vr_exp_validity_challenges, in_the_wild_exp}, necessitating robust anonymization and secure storage protocols~\cite{vr_exp_validity_challenges}.

\subsubsection{Experiment System Implementation}

\paragraph{Implementation Supports for Researchers without Technical Backgrounds}
\label{par:desirable-func-implementation-support}

VR research is expanding into multidisciplinary fields, such as psychology, social neuroscience, and healthcare~\cite{vr_across_fields, vr_healthcare, vr_rehabilitation}, yet current authoring platforms (e.g., Unity\footnote{\url{https://unity.com/}}, Unreal Engine\footnote{\url{https://www.unrealengine.com/}}) overwhelm domain experts who lack programming expertise~\cite{vr_across_fields, vr_healthcare}.
Implementation templates or low/no-code interfaces can lower this barrier and make VR practical for a wider range of researchers.

Given the presented diverse requirements, a comprehensive solution unifying them into a structured methodology would be promising for advancing VR research.

\begin{table*}[ht]
\centering
\caption{Comparison of VR/HCI Experimentation Tools and Platforms by Their Support for Experimentation Phases. \checkmark: Supported; $\triangle$: Partially or limitedly supported; -: Not supported.}
\label{tab:tools-platforms-comparison}
\begingroup
\resizebox{\textwidth}{!}{%
\begin{tabular}{@{} c c c c c c c c c c c c c @{}}
& \textbf{Sona} & \makecell[c]{\textbf{Jikken-}\\\textbf{Baito}} & \makecell[c]{\textbf{Microsoft}\\\textbf{Rocketbox}} & \textbf{HaRT} & \makecell[c]{\textbf{VRQues-}\\\textbf{tionnaire-}\\\textbf{Toolkit}} & \textbf{QuickVR} & \textbf{bmlTUX} & \textbf{UXF} & \textbf{USE} & \textbf{Ubiq-exp} & \makecell[c]{\textbf{Social}\\\textbf{VR-based}\\\textbf{attempts}} & \textbf{LUIDA} \\
\hline
\hline
\makecell[l]{\textbf{Large, Diverse}\\\textbf{User Pools}} & \checkmark & \checkmark & - & - & - & - & - & - & - & - & $\triangle$ & \checkmark \\
\midrule
\makecell[l]{\textbf{Access to}\\\textbf{VR Users}} & $\triangle$ & $\triangle$ & - & - & - & - & - & - & - & - & \checkmark & \checkmark \\
\midrule
\makecell[l]{\textbf{Multi-Stage}\\\textbf{Experiment}\\\textbf{Management}} & - & - & - & - & - & \checkmark & \checkmark & \checkmark & \checkmark & \checkmark & - & \checkmark \\
\midrule
\makecell[l]{\textbf{Avatar Control}} & - & - & \checkmark & - & - & \checkmark & - & - & - & \checkmark & - & \checkmark \\
\midrule
\makecell[l]{\textbf{Asynchronized}\\\textbf{Execution}} & - & - & - & - & - & - & - & - & - & \checkmark & - & \checkmark \\
\midrule
\makecell[l]{\textbf{Real-time}\\\textbf{Behavioral}\\\textbf{Tracking}} & - & - & - & - & - & - & - & \checkmark & \checkmark & \checkmark & - & \checkmark \\
\midrule
\makecell[l]{\textbf{In-VR}\\\textbf{Questionnaires}} & - & - & - & - & \checkmark & - & - & - & - & \checkmark & - & \checkmark \\
\midrule
\makecell[l]{\textbf{Customized}\\\textbf{data collection}} & - & - & - & \checkmark & \checkmark & \checkmark & \checkmark & \checkmark & \checkmark & \checkmark & $\triangle$ & \checkmark \\
\midrule
\makecell[l]{\textbf{Implementation}\\\textbf{Support}} & - & - & - & \checkmark & - & \checkmark & \checkmark & $\triangle$ & $\triangle$ & $\triangle$ & $\triangle$ & \checkmark \\
\midrule
\makecell[l]{\textbf{Hardware}\\\textbf{integration}} & - & - & - & - & - & \checkmark & - & - & \checkmark & $\triangle$ & - & \checkmark \\
\midrule
\end{tabular}
}
\endgroup
\end{table*}

\subsection{Comparison between Existing Tools and LUIDA}
\label{subsec:comparison-tools-and-luida}

Table~\ref{tab:tools-platforms-comparison} compares existing tools, previous social VR-based attempts, and LUIDA across each functionality.
While specialized tools remain fragmented and comprehensive frameworks require technical expertise or lack recruitment support, social VR attempts~\cite{recruit_on_vrchat} offer user access but limited efficiency and customizability.
LUIDA addresses most functionalities as a comprehensive and structured solution for remote VR studies.

The following listed the features and limitations of existing tools by each experimentation process, and introduce how LUIDA addresses them.

\subsubsection{Participant Recruitment}
Platforms like Jikken-baito~\cite{jikken_baito} and Sona~\cite{sona} handle only recruitment and lack dedicated VR-experienced participant pools.
Previous social VR attempts~\cite{recruit_on_vrchat} better access this user base but rely on inefficient, manual methods like inviting users in person within the platforms.
LUIDA streamlines this with a dedicated virtual recruitment world, enabling access to a broad, ecologically valid user pool.

\subsubsection{Experiment Execution}
Existing frameworks like QuickVR~\cite{quickvr}, bmlTUX~\cite{bmlTUX}, UXF~\cite{UXF}, USE~\cite{USE}, and Ubiq-exp~\cite{ubiq-exp} leverage multi-stage management (e.g., event sequences, session-block-trial models, state machines~\cite{state_machine}), which LUIDA also adopts.
In addition, Microsoft Rocketbox~\cite{rocketbox} provides a large library of avatar models, and QuickVR features support for handling avatar embodiment, encouraging LUIDA to incorporate avatar control mechanisms.
Furthermore, many existing tools lack inherent support for asynchronous execution.
While toolkits like Ubiq-exp can facilitate it, LUIDA similarly enables automated, large-scale parallel sessions without researcher oversight by leveraging metaverse platforms' capability to host multiple worlds.

\subsubsection{Data Collection}
LUIDA automates behavioral tracking and custom data collection, improving the manual approaches in social VR attempts and matching the capabilities of tools like HaRT~\cite{hart}, bmlTUX, UXF, USE, and Ubiq-exp, with an added web console to easily access results.
It also integrates in-VR questionnaires, a feature typically absent in existing frameworks except VRQuestionnaireToolkit~\cite{VRQuestionnaireToolkit} and Ubiq-exp.
For hardware integration, where QuickVR supports motion-capture and USE offers precise synchronization with external hardware, LUIDA provides a flexible alternative via the Open Sound Control (OSC) protocol\footnote{\url{https://ccrma.stanford.edu/groups/osc/index.html}}.

\paragraph{Experimental System Implementation}
While implementation is the primary function of frameworks like HaRT, QuickVR, bmlTUX, UXF, USE, and Ubiq-exp, they present significant barriers: UXF, USE, and Ubiq-exp have a high technical barrier requiring complex configuration files or expertise in Unity, coding, or networking; bmlTUX and HaRT have a more user-friendly graphical interface to configure the experiment flow, but bmlTUX is more restricted to factorial designs compared to the other frameworks, while HaRT is dedicated to hand redirection-related tasks.
LUIDA addresses this by providing an accessible, low-code Unity template that abstracts this complexity for a wider range of researchers.

In summary, LUIDA integrates the fragmented processes in VR experimentation, including participant recruitment, experiment execution, data collection, and system implementation.
This integration is expected to not only solve the complexity of experimental design and the difficulty of large-scale data collection, but also present a structured experimental paradigm that utilizes a metaverse environment for VR research.
Detailed designs of LUIDA are explained in later sections.

\subsubsection{Metaverse-Based Strategies for VR Experiments}
The integration of metaverse platforms with research methodologies has emerged as a critical approach to facilitate remote VR studies. Radiah~\etal positioned this approach as a key strategy for remote VR research, highlighting benefits such as high accessibility to existing user bases and reduced development overhead~\cite{remote_vr_exp_framework}. However, several challenges remain, including limitations in data collection methods, experimental control, and heterogeneity of participant devices, all of which can affect the internal validity of experiments~\cite{validity_reliability}.
While Saffo~\etal successfully validated the use of social VR for remote experiments, their methodology was limited to hosting sessions one by one and relied on inefficient manual procedures, such as taking pictures of results logged on a virtual wall~\cite{remote_exp_social_vr}.
This highlights the need for automated approaches that can parallelize sessions and data collection.

Saffo~\etal identified two strategies for conducting remote studies via social VR: partnering with existing platforms or building a research-dedicated platform from scratch.
The partnership model leverages established infrastructure and user bases but lacks complete customizability, while dedicated platforms offer full customization at significant development cost.
We decided to have LUIDA adopt a hybrid strategy, partnering with existing platforms while incorporating research-tailored features to provide both established infrastructure and enhanced customizability.
Meanwhile, we positioned LUIDA as a platform-agnostic design to ensure its core functionalities remain portable across different metaverse environments.

\begin{figure*}[ht]
    \centering
    \includegraphics[width=\linewidth]{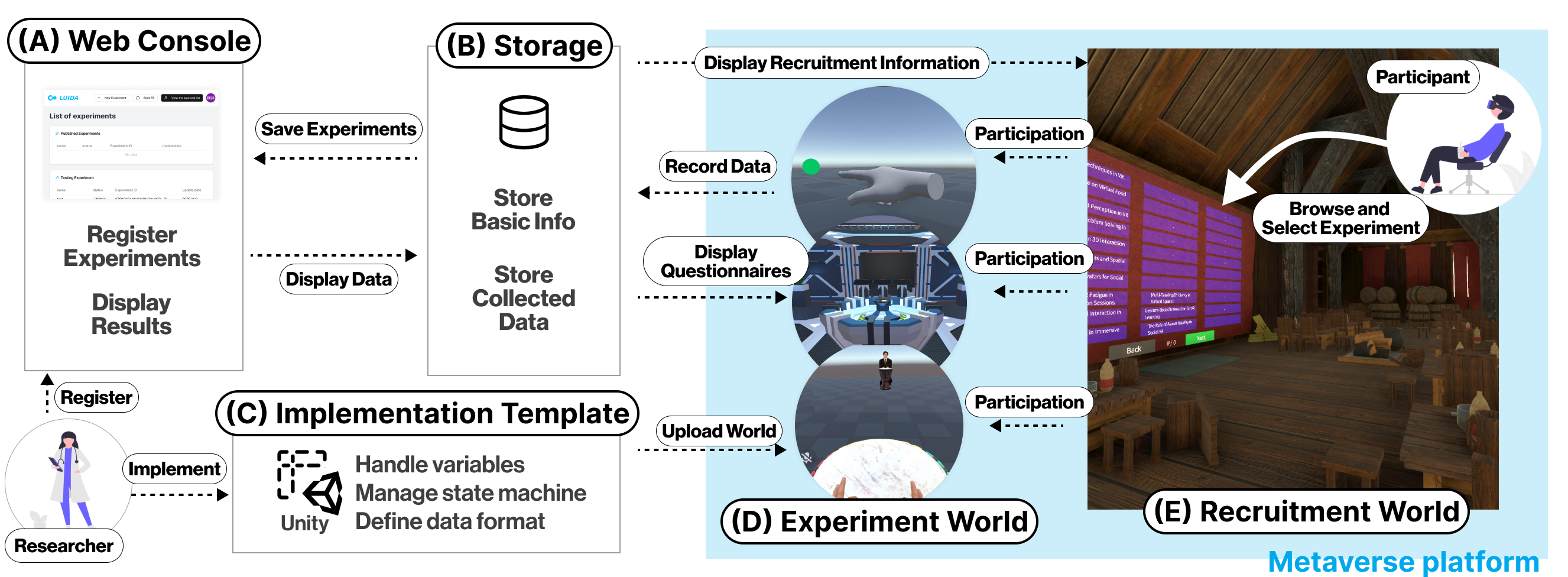}
    \caption{An overview of the LUIDA system architecture, consisting of five interconnected components. (A) Web Console: An interface for researchers to register experiments and visualize collected results. (B) Storage: A central database that stores basic experiment information registered via the Console and data collected during experiments. (C) Implementation Template: A Unity-based environment where researchers implement experimental logic, such as variable handling and state machines, to create the Experiment World. (D) Experiment World: The virtual environment generated from the template where the experiment takes place; it automatically displays questionnaires and records participant data to Storage. (E) Recruitment World: The entry point on the metaverse platform where participants browse recruitment information fetched from Storage and access specific Experiment Worlds.}
    \label{fig:system-overview}
\end{figure*}

\section{LUIDA: Framework Proposal}\label{sec:overview}
This section introduces the proposed framework of LUIDA, including the overall system architecture and the expected user interaction flow for researchers and participants.

\subsection{System Architecture}
\label{subsec:architecture}
Figure~\ref{fig:system-overview} illustrates the overall architecture of LUIDA, which consists of five main components:
\begin{enumerate}
    \item \textbf{Recruitment World:} The virtual environment where participants can browse and select available experiments.
    \item \textbf{Experiment World:} The virtual environment in which experiments are carried out, including questionnaire displays and data collection functions.
    \item \textbf{Implementation Template:} A Unity-based (as Unity is the standard world editor for platforms like VRChat or Cluster) virtual space editor that provides no-code interfaces for researchers to construct experimental systems and provides interfaces for variable management, state machine control, and format definition for data collection.
    \item \textbf{Web Console:} A web interface for experiment registration, questionnaire configuration, and data visualization.
    \item \textbf{Storage:} A database system for storing experimental information and collected data.
\end{enumerate}

These components interact through the following data flows:

\begin{itemize}
\setlength{\parskip}{0.1cm}
\item \textbf{Web Console and Storage:}  The Web Console uses Storage to register experiments and retrieve data.
\item \textbf{Recruitment World and Storage:} The Recruitment World queries Storage to update its public bulletin board of available experiments.
\item \textbf{Experiment World Generation:} When a participant joins, the system instantiates a unique Experiment World based on a world the researcher created with the Implementation Template.
\item \textbf{Experiment World Execution, Configuration in the Implementation Template, and Storage:} During an experiment, the Experiment World automatically applies the parameters, state machine, and data formats defined in its Implementation Template.
\item \textbf{Questionnaire Generation:} The Experiment World displays questionnaires by loading items from Storage based on an ID set in the Implementation Template.
\item \textbf{Data Collection Flow:} All data from the Experiment World is saved to Storage and can be accessed by researchers through the Web Console.
\end{itemize}

Note that the name ``LUIDA'' is derived from ``Luida's Bar,'' a location in the popular role-playing game series ``Dragon Quest\footnote{\url{https://www.dragonquest.jp/}}'', and reflects our expectation that users will participate in experiments and receive rewards in a manner similar to accepting quests in games.

\subsection{User Interaction Flow}

\subsubsection{Participant Workflow} 
The participant workflow in LUIDA follows these steps (Fig.~\ref{fig:participation-experience}):
\begin{figure*}[t]
    \centering
    \includegraphics[width=\linewidth]{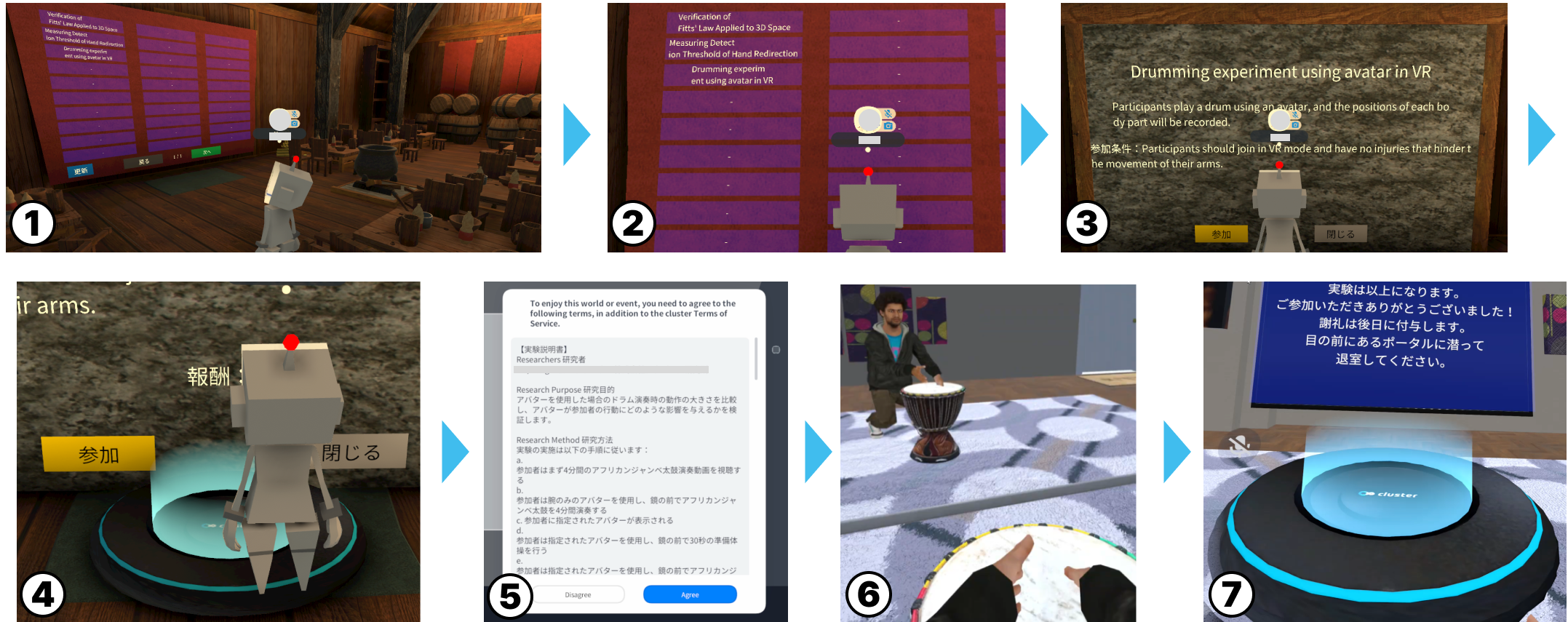}
    \caption{Participant workflow within LUIDA. (1) Entry: Participants enter the shared Recruitment World. (2) Browsing: They view the bulletin board listing available studies. (3) Selection: Selecting an experiment displays its details and requirements. (4) Transition: Pressing a "Join" button spawns a portal to the specific Experiment World. (5) Consent: Upon arrival, participants must review and agree to the consent form. (6) Execution: Participants perform the experimental tasks (e.g., drumming; see Figs. \ref{fig:hand-redirection-exp-procedure}, \ref{fig:drumming-avatar-exp-procedure}, and \ref{fig:fitts-3d-exp-procedure} for details). (7) Completion: A final portal returns them to the Recruitment World to receive rewards.}
    \label{fig:participation-experience}
\end{figure*}

\begin{enumerate}
    \item \textbf{Entering the Recruitment World:} Participants enter the Recruitment World to browse a bulletin board listing available experiments with information such as requirements and rewards.
    \item \textbf{Experiment Selection and Participation:} After selecting an experiment, a portal appears for participants to enter the Experiment World, where they provide consent and begin the tasks.
    \item \textbf{Task Execution and Data Collection:} Participants answer in-VR questionnaires or perform tasks following the instructions. During this process, positional data from VR headsets is automatically recorded and applied to avatars' full-body movements using inverse kinematics.
    \item \textbf{Experiment Completion and Reward:} After completion, participants receive rewards and return to the Recruitment World, where they can choose to participate in another experiment or leave the metaverse environment.
\end{enumerate}

\subsubsection{Researcher Workflow}
Researchers using LUIDA to design and conduct experiments follow these steps:

\begin{enumerate}
\item \textbf{Experiment Registration in Web Console:} Researchers access the Web Console to register experiment information (title, description, participation requirements, rewards, consent form content, etc.) and questionnaire content. Questionnaire identifiers set in the Implementation Template are linked to the actual questionnaire items set within the Web Console information.
\item \textbf{Implementation Template Download and Experiment World Development:} Researchers download the Unity-based Implementation Template to build the Experiment World. The template offers interfaces to link questionnaires (detailed in Sec.~\ref{subsec:overview:question}), define data formats (detailed in Sec.~\ref{subsec:overview:tracking}), and automate the experimental flow using a state machine and variable management (detailed in Sec.~\ref{subsec:overview:automated-execution}).
\item \textbf{Experiment World Upload and Publication:} Researchers upload the completed Experiment World, making it accessible from the Recruitment World. Researchers can then switch its public visibility on the Web Console (e.g., hiding it once enough data is collected).
\item \textbf{Data Collection and Analysis:} As experiments are completed, data is automatically saved to Storage. Researchers can monitor and download results for analysis from the Web Console.
\end{enumerate}

\section{LUIDA: Core Features and Mechanisms}\label{sec:core-features}
LUIDA integrates a variety of features and mechanisms for the comprehensive management of VR experiments leveraging the metaverse environment.
In this section, we describe the key features of LUIDA, which align with the critical functionalities for VR experiments identified in Section~\ref{subsec:related:desirable-func}.

\subsection{In-Situ \& Automated Participant Recruitment}
\label{subsec:overview:automated-recruitment}
This feature addresses the desirable functionalities presented in Sec.~\ref{par:desirable-func-larger_and_more_diverse_user_pools} and Sec.~\ref{par:desirable-func-vr-users}.
Many metaverse platforms host multiple, individual virtual environments (called `Worlds' in VRChat or Cluster\footnote{\url{https://cluster.mu/}}, `Rooms' in Rec Room\footnote{\url{https://recroom.com/}}, or `Regions' in Second Life\footnote{\url{https://secondlife.com/}}) where users gather and socialize.
Leveraging this, LUIDA features a dedicated Recruitment World (as proposed in Sec.~\ref{subsec:architecture}) with a tavern-like appearance to encourage gathering and prompt in-situ recruitment.
Inside, a bulletin board automates the recruitment process by listing all available studies, allowing users to browse and join at will, eliminating the need for researchers to actively seek participants.

\subsection{Experiment Execution}
\subsubsection{Asynchronized and Parallel Execution}
\label{subsec:overview:parallel-execution}
Addressing the desirable functionality presented in Sec.~\ref{par:desirable-func-async-exec},
LUIDA utilizes the metaverse platforms' world/room generation capabilities, automatically generating a unique, isolated instance for each session using the Experiment World as a template.
This enables massive parallel sessions and accelerates data collection without researcher oversight.
To ensure data quality, the system prevents re-entry into completed experiments and can redirect non-VR users if an experiment is VR-exclusive.

\subsubsection{Avatar Control and Management}\label{subsec:overview:avatar}
Addressing the desirable functionality presented in Sec.~\ref{par:desirable-func-avatar},
LUIDA supports avatar-related studies with a management mechanism allowing researchers to upload and configure avatars via the Web Console. 
Researchers can treat avatar characteristics (appearance, movements, etc.) as experimental variables.
In addition, LUIDA's state machine functionality helps ensure that participants acclimate to their virtual bodies, critical for establishing proper embodiment before trials begin.

\subsection{Data collection}\label{subsec:overview:tracking}

\subsubsection{In-VR Questionnaire}\label{subsec:overview:question}
\begin{figure}[ht]
    \centering
    \includegraphics[width=\linewidth]{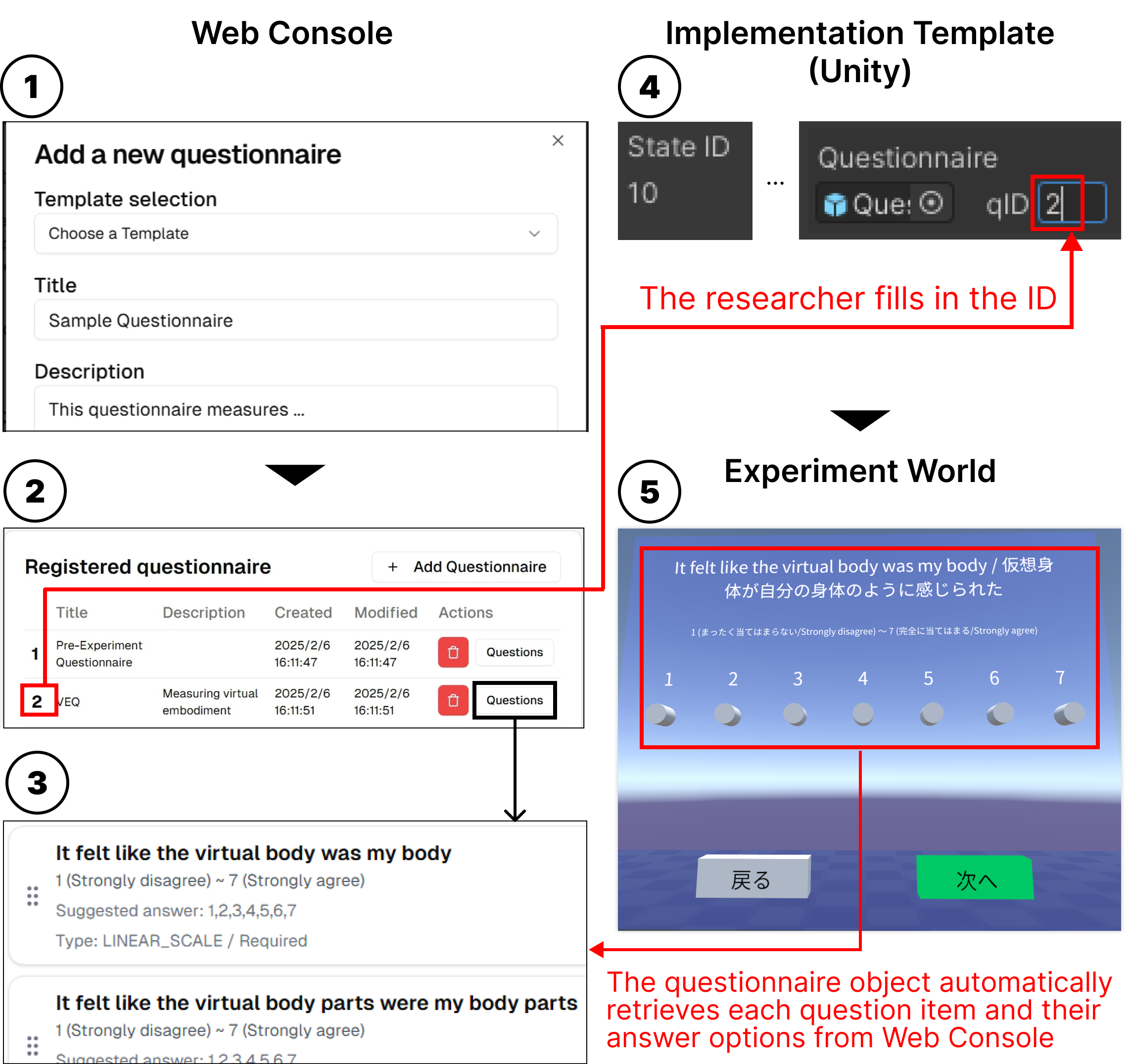}
    \caption{Configuration procedure for automated questionnaire generation: (1) In the LUIDA Web Console, users register a questionnaire using a form. (2) From the list of registered questionnaires, they click the ``Questions'' button to edit the question items (3). (4) In the Implementation Template (the screenshot captures the actual interface within the Implementation Template of LUIDA's prototype), the corresponding questionnaire ID is set within the state where the questionnaire should appear. (5) During the experiment, the Experiment World automatically generates questionnaire objects based on the registered question items (3).
    }
    \label{fig:link_questionnaire}
\end{figure}

Addressing the desirable functionality presented in Sec.~\ref{par:desirable-func-questionnaire},
LUIDA implements in-VR questionnaires without requiring participants to remove their HMDs.
As illustrated in Fig.~\ref{fig:link_questionnaire}, researchers register questions and response formats (e.g., Likert scale, multiple-choice) in the Web Console, then link to the questionnaire via its ID in the Implementation Template.
LUIDA then automatically generates the in-VR questionnaire during the experiment by loading items from Storage, handling all UI and layout.

\subsubsection{Continuous User Tracking}
Addressing the desirable functionality presented in Sec.~\ref{par:desirable-func-behavioral-tracking},
LUIDA automatically records position and rotation data from HMDs, controllers, and other tracking devices frame-by-frame throughout the experiment.
This enables quantitative evaluation of subtle behavioral patterns such as proxemics~\cite{Yaremych2019TracingPB}.

\subsubsection{Customizable Data Collection} Beyond standard tracking data, LUIDA allows researchers to define custom data formats within a dedicated file in the Implementation Template to record experiment-specific variables.
For example, task scores can be logged trial-by-trial and uploaded upon completion.
Once an experiment is running, data is collected automatically in the configured format, and researchers can later download them from the Web Console.

\subsubsection{Integration with External Systems}
Addressing the desirable functionality presented in Sec.~\ref{par:desirable-func-hardware},
LUIDA integrates with external software or hardware via the host metaverse platform's data streaming features, such as the OSC protocol supported by VRChat and Cluster.
This capability broadens LUIDA's utility to experiments requiring real-time physiological data, such as EEG or eye-tracking.

\subsubsection{Data Integrity and Privacy}
\label{par:data-integrity-privacy}
This feature addresses the desirable functionality presented in Sec.~\ref{par:desirable-func-data-integrity}.
To maintain both data integrity and participant privacy, all collected data are assigned unique identifiers that ensure anonymity while preserving data consistency.
Timestamps are also attached to establish a reliable chronological record for data integrity.

\subsection{Configuration in Implementation Template to Support Automated Execution}
\label{subsec:overview:automated-execution}

Basic operations like placing virtual objects can be delegated to Unity's native functionalities, while LUIDA focuses on supporting implementation and automation of experiment flows without requiring advanced manupulations or coding.
LUIDA's Implementation Template provides no-code interfaces for a state machine (controlling event sequences) and automatic trial generation based on configured variables.
These features addresses the desirable functionality presented in Sec.~\ref{par:desirable-func-multi-stage}, with the no-code interface addressing Sec.~\ref{par:desirable-func-implementation-support}.
Details are described in the subsequent paragraphs.

\begin{figure}[ht]
    \centering
    \includegraphics[width=\linewidth]{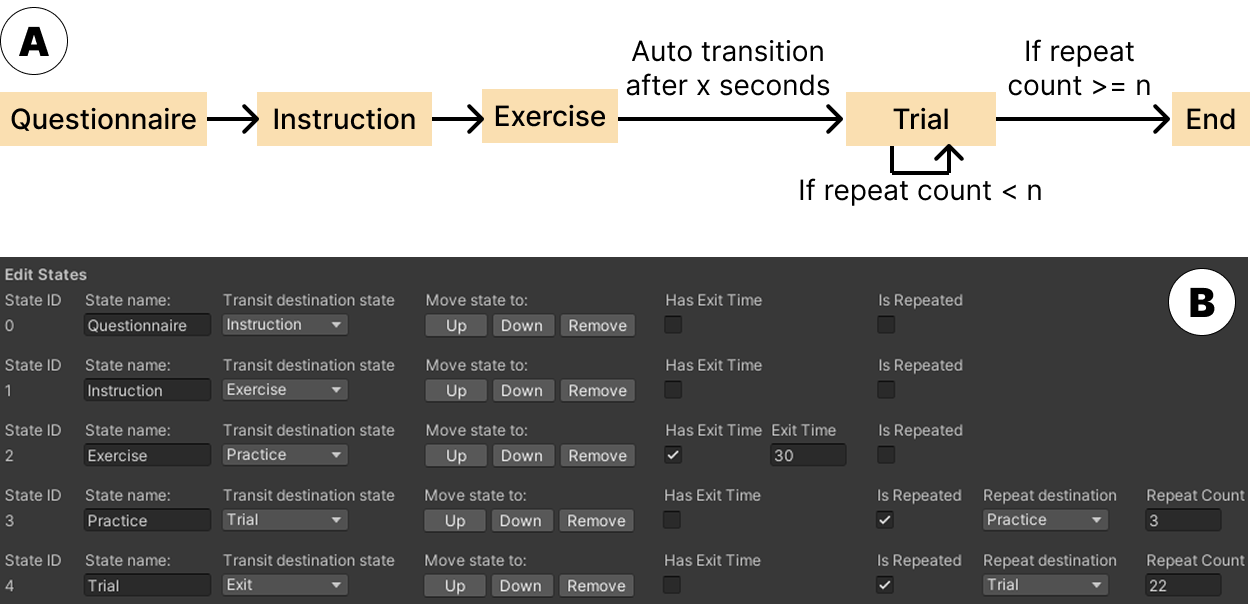}
    \caption{Configuration of the state machine that manages the entire experiment flow. (A) Example setup illustrating state transitions; arrows without text indicate transitions to be triggered manually (e.g., when participants click a button or complete a questionnaire). (B) Interface for configuring these settings within the Implementation Template of LUIDA’s prototype.}
    \label{fig:state-transition}
\end{figure}

\begin{figure}[ht]
    \centering
    \includegraphics[width=0.5\linewidth]{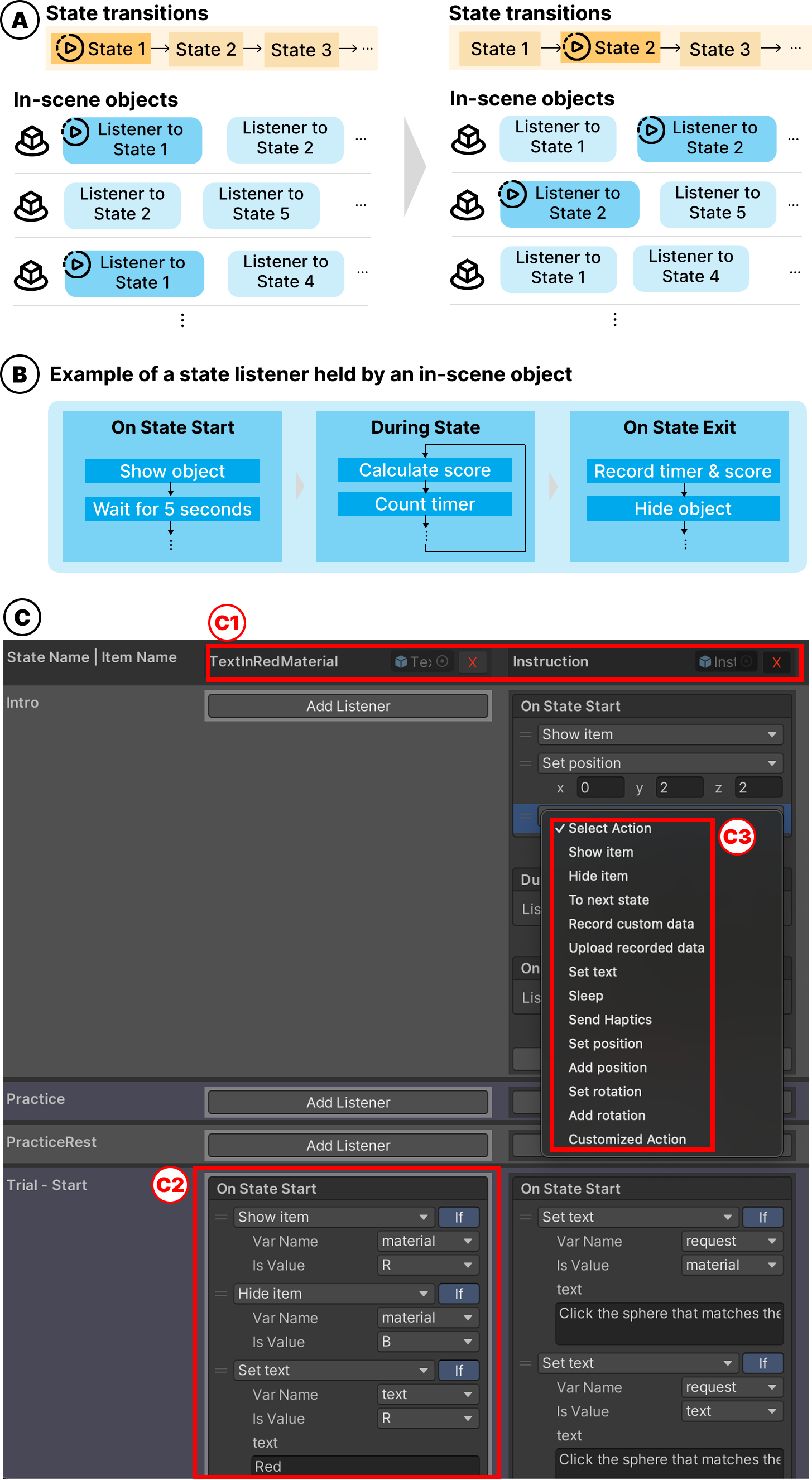}
    \caption{Configuration of objects listening to state transitions utilizing the Observer pattern. (A) Each in-scene object holds multiple state listeners, each monitoring a specific state. When the monitored state is activated, the corresponding listener is triggered to execute its assigned actions. (B) An example of a state listener attached to an in-scene object, showing registered actions to be performed at each lifecycle point: upon entering the monitored state (\textit{On State Start}), every frame during the state (\textit{During State}), and upon exiting the state (\textit{On State Exit}). (C) The configuration interface within the LUIDA prototype’s Implementation Template, comprising: (C1) a list of in-scene objects; (C2) a state listener with actions registered at the \textit{On State Start} lifecycle point; and (C3) a dropdown menu of available actions to register.}
    \label{fig:state-and-object-settings}
\end{figure}

\subsubsection{Experiment Flow Management via State Machine}\label{subsec:state-machine}
LUIDA's Implementation Template employs a centralized finite state machine~\cite{state_machine,unity_state_pattern} to support controlling complex experimental flows, such as implementing immersive guidance by dynamically relocating instruction panels, displaying virtual mirrors for avatar acclimation, switching object visibility and sending haptic feedback upon interaction, triggering questionnaires or data collection at specific timings, and so on.

Via a no-code interface in the Implementation Template, researchers define states and their transitions, with options to control any state's duration or repetition for repeated tasks (Fig.~\ref{fig:state-transition}).
In addition, inspired by QuickVR's workflow system~\cite{quickvr}, LUIDA adopts the observer pattern~\cite{observer_pattern}, allowing researchers to attach ``state listeners'' to in-scene objects through a graphical interface, on which state listeners can be configured to trigger actions at each state's lifecycle point: \textit{On State Start}, \textit{During State}, and \textit{On State Exit}.
Available actions include changing visibility or position of a player/object, setting text, sending haptic feedback to VR controllers, saving data, sending an OSC message (to communicate with external hardware), and more.
Additionally, a ``Customized Action'' code block allows researchers with programming expertise.

\subsubsection{Variable Configuration and Automated Trial Generation}\label{subsec:variables-config}
LUIDA's Implementation Template enables configuration of experimental variables and automatically generates trials based on these variables, reducing human error and preparation time.
This feature benefits factorial designs, ensuring participants experience all factor combinations in either randomized order (to counteract order effects) or fixed order (for example, to analyze learning effects).

\begin{figure}[ht]
    \centering
    \includegraphics[width=0.8\linewidth]{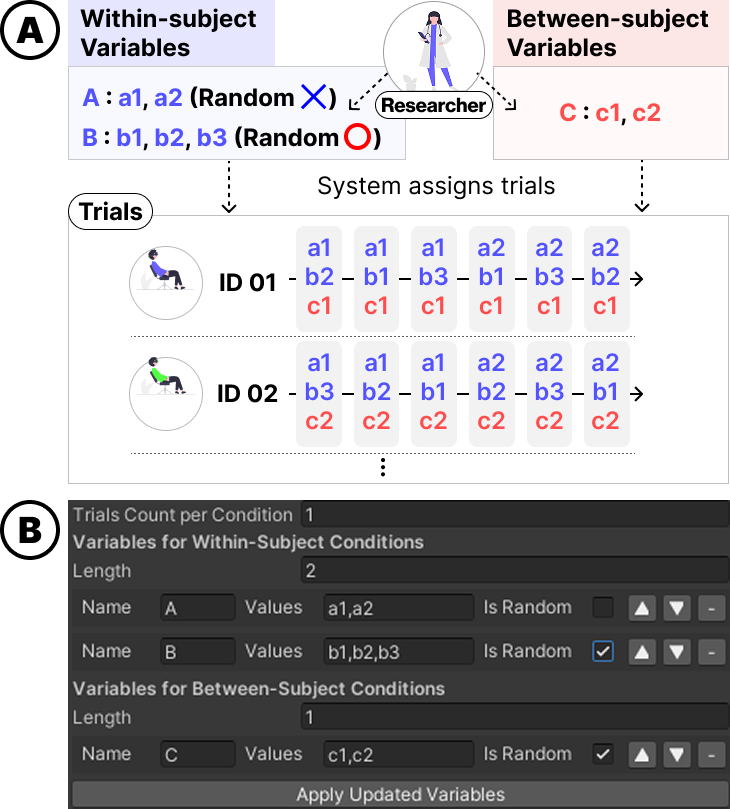}
    \caption{Variable configuration functionality of the Implementation Template: (A) An example of variable setup and the corresponding generated trials based on the configuration; (B) The actual interface within the Implementation Template (a Unity project) of LUIDA's prototype.}
    \label{fig:variable-settings}
\end{figure}

Through a no-code interface (Fig.~\ref{fig:variable-settings}), researchers can visually define both within- and between-subject variables with their possible values, trial repetitions, and randomization options.
Then, during experiment execution, values for between-subject variables are randomly assigned.
Next, the state machine introduced in Sec.~\ref{subsec:state-machine} includes trial-dedicated states, which iterate through all unique combinations of within-subject variables $\times$ the number of trial repetitions.
Actions within trials are configurable as described in Sec.~\ref{subsec:state-machine}, while values of both within- and between-subject variables are usable as conditions to control whether to trigger an action.

\section{Implementation}
This section describes the practical implementation of the LUIDA system architecture described in Sec.~\ref{sec:overview}.

\subsection{Metaverse Platform Selection}
\subsubsection{Cluster Platform Selection Rationale}
\begin{figure}[ht]
    \centering
    \includegraphics[width=0.9\linewidth]{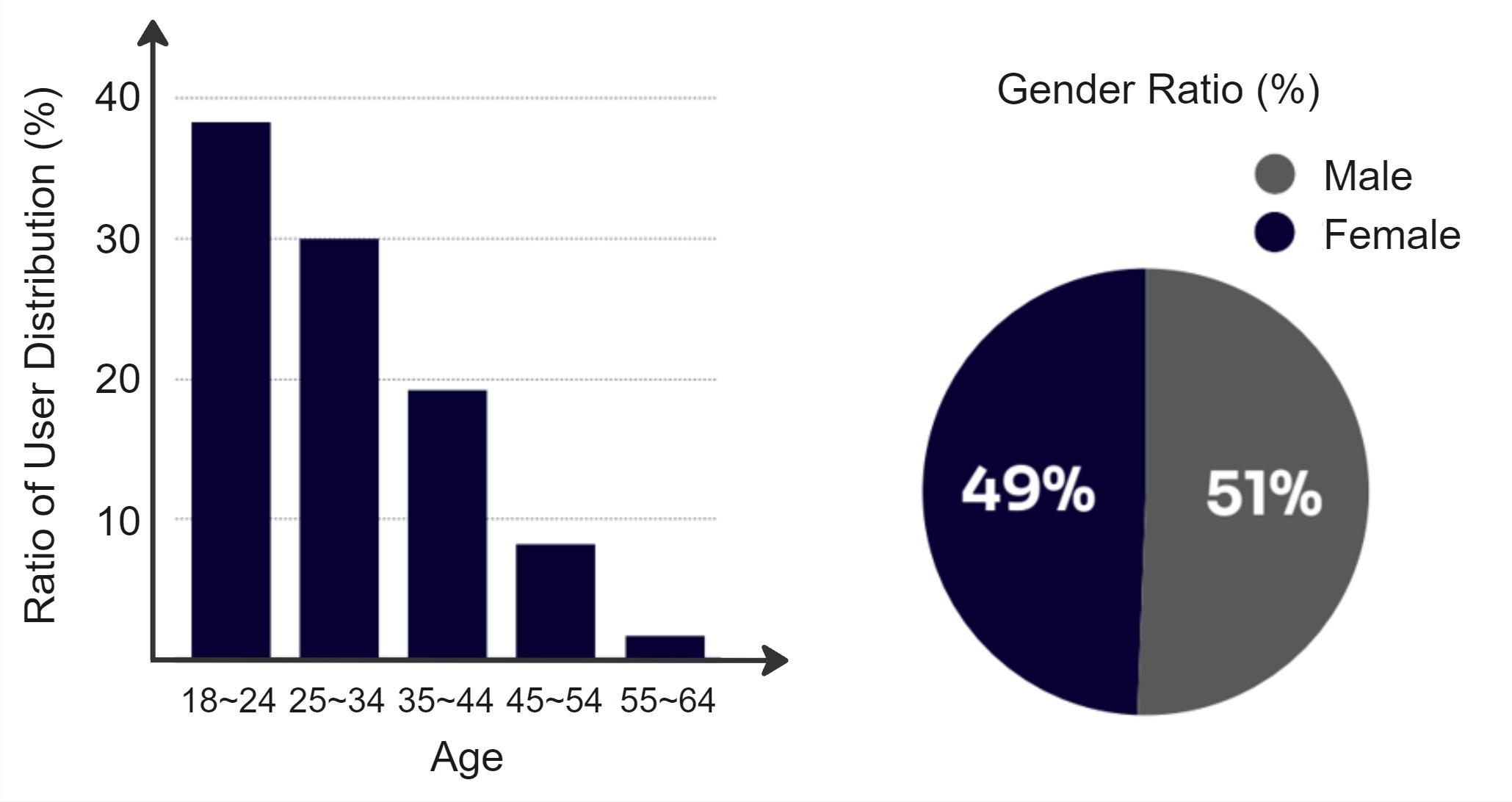}
    \caption{Demographic information of Cluster's users across different age groups and with balanced gender ratio.}
    \label{fig:user_distribution}
\end{figure}

We chose the metaverse platform Cluster for the LUIDA prototype for three key reasons:
First, its diverse user base (spanning various age groups with a balanced gender ratio, shown in Fig.~\ref{fig:user_distribution}) potentially increases the generalizability of research findings.
Second, its content creation tools: the Unity-based Cluster Creator Kit (CCK) for no-code development and ClusterScript for advanced scripting ensure flexibility for researchers with different expertise levels.
Third, Cluster supports various SteamVR\footnote{\url{https://store.steampowered.com/steamvr}}-compatible VR headsets (Meta Quest, HTC Vive, Valve Index, Pico, etc.), increasing participant accessibility.
While this paper focuses on VR experiments, Cluster also supports PC and mobile devices, extending LUIDA's potential to broader HCI experiments.

We identified platform-specific constraints of Cluster, such as lacking native finger tracking and pass-through support, and allow avatar assignment only upon world entry, posing a challenge for experiments that require dynamic avatar switching.
We have implemented a special solution to this limitation, as described in Sec.~\ref{subsec:impl:avatar-control}.
Despite these limitations, Cluster's large user base, robust creation tools, and broad device compatibility make it a suitable foundation for LUIDA's prototype.

\subsubsection{Platform-Agnostic Constraints and Mitigation Strategies}
\label{subsubsec:implementation-constraints-all-platforms}
LUIDA also faces constraints inherent to all metaverse platforms.
For instance, network latency can challenge experimental timing precision for cognitive experiments requiring millisecond-level synchronization.
LUIDA mitigates this with timestamp logging to help researchers identify delayed data.

LUIDA's scalability depends on the host platform's capacity for multiple world instances.
Given that platforms like VRChat maintain around 50,000 concurrent users in the year of 2025~\cite{vrchat-steamdb}), we did not consider this urgent, though future work could include formal stress tests.

\subsection{Component Implementation}
This subsection describes how we actually implemented LUIDA's core features.

\subsubsection{Recruitment World}
The Recruitment World is implemented as a public space accessible to all Cluster users.
This world features a bulletin board interface that displays all available experiments.
Its role-playing game-like design (a tavern-like appearance) reduces psychological barriers to participation.

\subsubsection{Experiment World}
Leveraging Cluster's automatic instance generation, when participant count reaches the configured limit (default: 1), the system automatically generates new instances for additional participants, allowing parallelized execution.
Experiment Worlds execute automatically according to researchers' state machine, variable, questionnaire, and data collection configurations within the Implementation Template.

\subsubsection{Implementation Template}
The Implementation Template was created as a Unity project template using CCK.
Interfaces for key functionalities, including questionnaire ID configuration (Fig.~\ref{fig:link_questionnaire}.4), state machine management (Fig.~\ref{fig:state-transition}B and ~\ref{fig:state-and-object-settings}C), experimental variable manipulation (Fig.~\ref{fig:variable-settings}B), and data collection format specification were developed in C\# within the Unity editor.
Configurations made through these interfaces are reflected in CCK and ClusterScript, eliminating researchers' need to write C\# code themselves.
Once uploaded as Experiment Worlds, the execution of these template-based configurations are fully delegated to CCK and ClusterScript.

\subsubsection{Web Console}
The Web Console, deployed on Vercel\footnote{\url{https://vercel.com/}} (screenshots in Supplementary Material  Sec. 1) provides the following functionality:
experiment management (register, edit, publish/unpublish experiments); questionnaire configuration (edit questions and response options); avatar settings; and data visualization and download.

\subsubsection{Storage} 
Collected data is stored on AWS S3\footnote{\url{https://aws.amazon.com/s3/}} with unique, non-traceable identifiers ensuring privacy.

\subsection{Dynamic Avatar Switching}\label{subsec:impl:avatar-control}
To overcome Cluster's limitation of assigning avatars only upon world entry, LUIDA allows researchers to register multiple worlds under a single experiment; each can be configured with a different avatar.
Then, redirecting participants through portals enables dynamic avatar switching.

\section{Evaluation Design}
\label{sec:exp-to-replicate}
\subsection{Research Questions}
The following research questions (RQs) were formulated regarding LUIDA's efficacy for experimentation:

\begin{description}[leftmargin=0pt,labelindent=1pt,itemsep=1pt]
    \item[RQ1] How does LUIDA benefit researchers conducting VR experiments?
    \item[RQ2] Can LUIDA accommodate a variety of VR experiments?
    \item[RQ3] How does LUIDA promote the participation of participants in experiments?
    \item[RQ4] Does LUIDA maintain experimental validity compared to traditional laboratory experiments?
\end{description}

We addressed these RQs through two studies.
Before conducting them, we sought to select target research domains to validate whether LUIDA could accommodate a wide variety of VR experiments.
Viewing the topics listed by major VR conferences such as IEEE VR~\cite{ieeevr-topics}, we observed that, excluding those focusing on infrastructure (e.g., ``XR technology infrastructure''), the remaining user-study-oriented topics can be broadly classified into three domains: perception (e.g., ``Cybersickness'', ``Multimodal/cross-modal interaction and perception''), cognition (e.g., ``Embodied agents, virtual humans and (self-)avatars'', ``User experience and usability''), and user interfaces (e.g., ``3D user interfaces'', ``Locomotion and navigation'').
For each domain, we chose a representative experiment for later validations: a hand redirection study by Zenner~\etal (Sec.~\ref{sec:exp-to-replicate-hand}) for perception, a Proteus effect study by Kilteni~\etal (Sec.~\ref{sec:exp-to-replicate-proteus}) for cognition, and a 3D Fitts' Law study by Clark~\etal (Sec.~\ref{sec:3d-fitts-law-intro}) for user interfaces.

Then, in the first study, we evaluated LUIDA's benefits and versatility (RQ1, RQ2) by inviting VR researchers to replicate the one of the three chosen experiments on LUIDA and collecting feedback from them (Sec.~\ref{sec:user-study-with-researchers}).
For the second study, we replicated the three chosen experiments on LUIDA and make them public to Cluster users.
The online participation and collected data allowed us to examine participant engagement and experimental validity (RQ3, RQ4) (Sec.~\ref{sec:user-study-online}).
All experiments were approved by the Cluster Metaverse Lab Ethics Committee (2024-003).

The remainder of this section describes in detail the experiments selected for replication.

\subsection{Measuring Detection Threshold of Hand Redirection}
\label{sec:exp-to-replicate-hand}
Hand redirection is a technique introducing discrepancies between real and virtual hand movements, often to enable interaction in a larger virtual space than the physical one affords.
We replicated the ``gain warp without distraction'' condition from Zenner~\etal, where participants repeatedly performed a reaching task and reported if their virtual hand felt faster or slower than their real hand, and a detection threshold (DT) was calculated from their responses~\cite{dt_hand_redirection}.

\begin{figure}[h]
    \centering
    \includegraphics[width=0.8\linewidth]{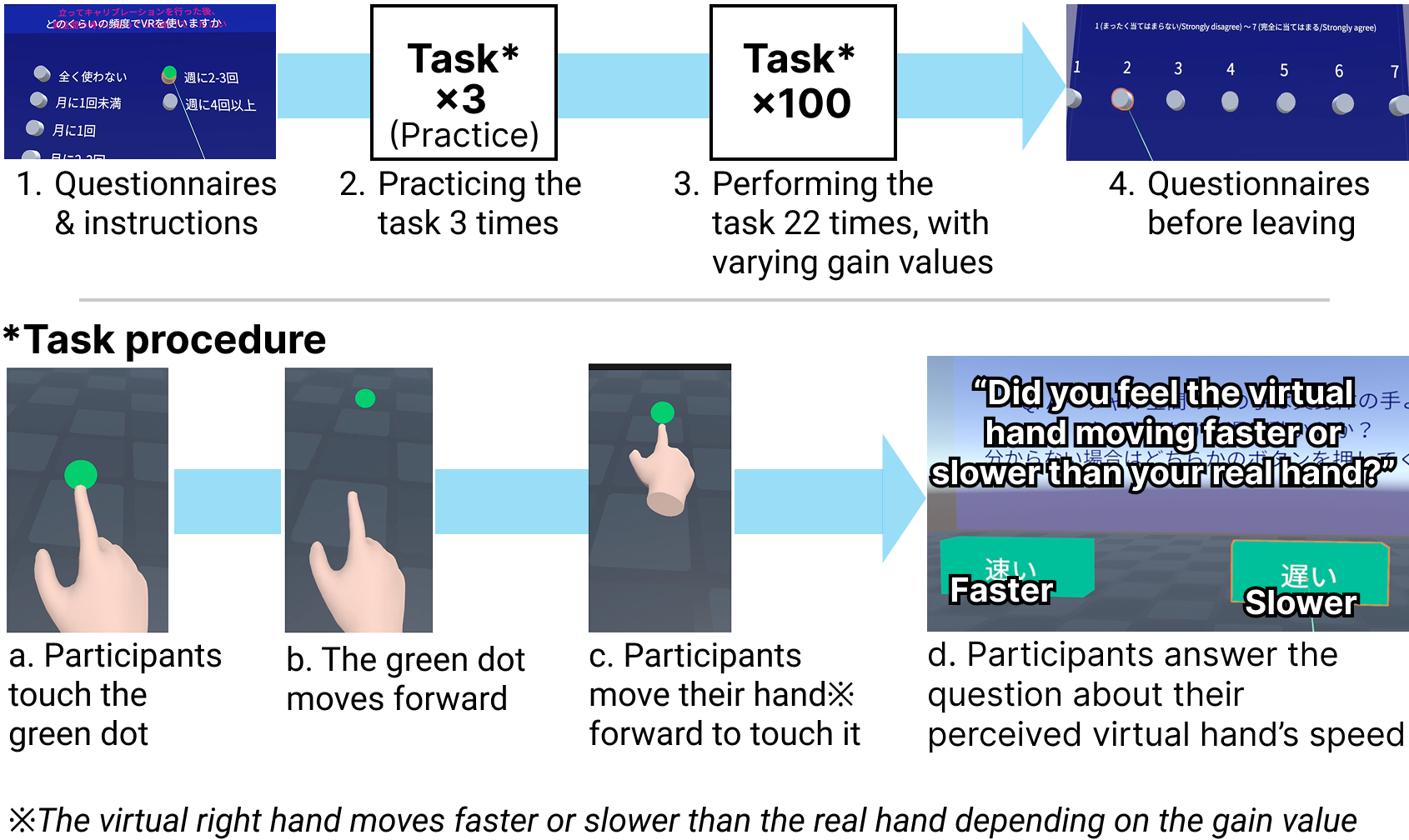}
    \caption{Procedure of the replicated hand redirection experiment.}
    \label{fig:hand-redirection-exp-procedure}
\end{figure}

The within-subjects procedure was outlined in Fig.~\ref{fig:hand-redirection-exp-procedure} and explained as follows:
\begin{enumerate}
    \item Participants joined the Experiment World and completed a questionnaire (demographics, VR experience).
    \item Instructions for the reaching task were provided.
    \item Practice session: One trial each for gain values of 0, 1.25, and 0.75.
    \item Two trials for each of 11 gain values (from 0.75 to 1.25) presented in a randomized order.
    \item Participants completed post-experiment questionnaires (Sense of Embodiment, Presence).
\end{enumerate}

\subsection{Proteus Effect during drumming in VR}
\label{sec:exp-to-replicate-proteus}

\begin{figure*}[t]
    \centering
    \includegraphics[width=1\linewidth]{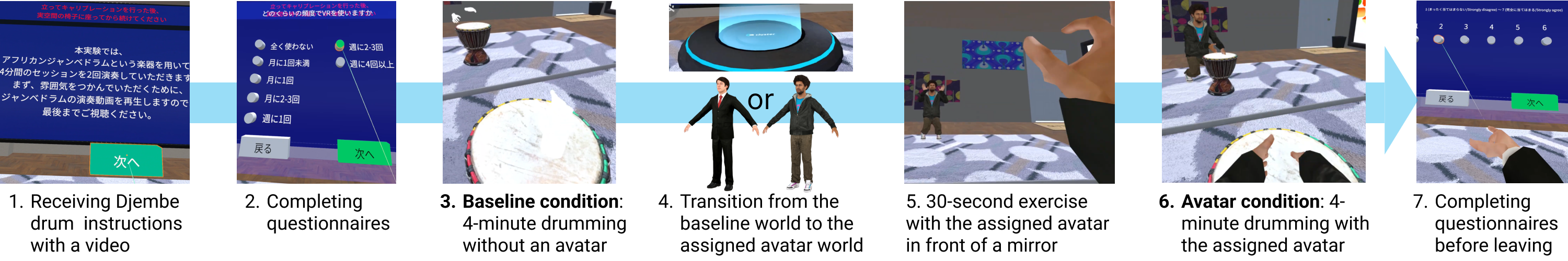}
    \caption{Procedure of the replicated drumming avatar experiment.}
    \label{fig:drumming-avatar-exp-procedure}
\end{figure*}

In Kilteni~\etal's study~\cite{drumming_proteus}, participants played an African djembe drum while using a Formal
Light-skinned (FL) avatar in a suit or a Casual Dark-
skinned (CD) avatar in casual clothing.
Performance was measured by analyzing body part positions during drumming to assess the cognitive effects induced by the avatar's appearance through behavioral changes.
We replicated this experiment using a between-subjects design with the following procedure (also outlined in Fig.~\ref{fig:drumming-avatar-exp-procedure}):

\begin{enumerate}
    \item Participants received instructions on playing a virtual Djembe drum.
    \item \textbf{Baseline}: Participants drummed for 4 minutes using a transparent avatar and monochrome hands while their HMD and controller positions were recorded.
    \item Participants were moved to a new world where they were assigned either a FL or CD avatar.
    \item Participants performed warm-up exercises in front of a mirror for 30 seconds to adjust to their new avatar.
    \item \textbf{Avatar condition}: Participants drummed for another 4 minutes with their assigned avatar, with movement data again being recorded.
    \item Participants completed a questionnaire on Sense of Embodiment.
\end{enumerate}

\subsection{Extending Fitts' Law to 3D Environment}
\label{sec:3d-fitts-law-intro}

\begin{figure}[h]
    \centering
    \includegraphics[width=0.8\linewidth]{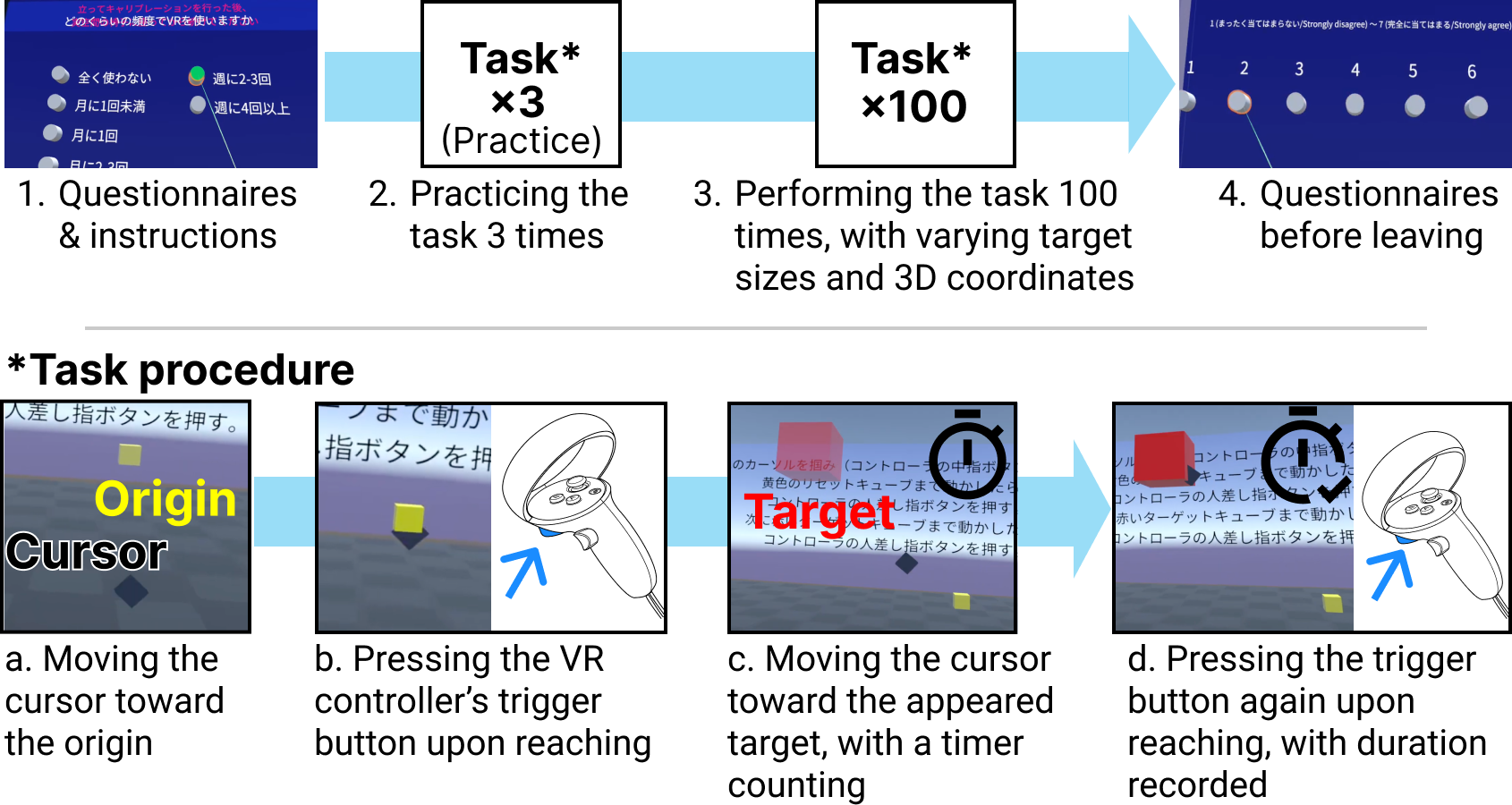}
    \caption{Procedure of the replicated 3D Fitts' Law experiment.}
    \label{fig:fitts-3d-exp-procedure}
\end{figure}

Fitts' Law is a model that predicts movement time to a target based on its distance and size.
Clark~\etal extended this law to 3D virtual environments by having participants move a virtual object to targets of varying sizes and 3D positions.
They also compared their model with the original Fitts' Law and other previous works of similar 3D extensions~\cite{fitts_original, fitts_murata_iwase, fitts_cha_myung, fitts_machuca}.
We replicated the experiment in a within-subjects design with the following procedure (also shown in Fig.~\ref{fig:fitts-3d-exp-procedure}): 

\begin{enumerate}
    \item Participants completed a pre-experiment questionnaire (demographics, VR experience).
    \item Participants received task instructions.
    \item Participants practiced the task three times.
    \item Main session: Participants performed 100 trials with varying target sizes and 3D coordinates.
    \item Participants completed a post-experiment questionnaire on Presence.
\end{enumerate}

\section{Usability Evaluation with VR Researchers}\label{sec:user-study-with-researchers}
We conducted a usability evaluation with VR researchers using LUIDA to investigate RQ1 and RQ2.

\subsection{Experiment Setup}
\subsubsection{Participants and Procedure}
We recruited 6 researchers with experience in VR experiments (age = 23.83 $\pm$ 1.83 years) using the kinship method.
Each was assigned one of the three experimental designs from Sec.~\ref{sec:exp-to-replicate} that best matched their expertise and was tasked with implementing it using LUIDA, rewarded with a 15,000 JPY ($\approx$ 100 USD) Amazon gift card.

The study was conducted asynchronously, with participants completing the following steps within a one-week timeframe:
\begin{enumerate}
    \item Read an introduction and manual of LUIDA.
    \item Download the Implementation Template and implement the experiment according to the provided manual.
    \item Register and publish the implemented experimental system via the Web Console of LUIDA.
\end{enumerate}

After implementation, each researcher tested their own experiment by accessing it from the Recruitment World, running the tasks, and reviewing the collected data on the Web Console.
Finally, they provided feedback on LUIDA's benefits and limitations through questionnaires and an interview.

\subsubsection{Evaluation Metrics}
We used the System Usability Scale (SUS)~\cite{sus} for usability and NASA-TLX~\cite{nasatlx} for workload.
These quantitative metrics were collected twice: once for the Implementation Template specifically and again for the overall LUIDA workflow (from registration to data review).
We also conducted interviews to gather qualitative insights into their ratings, perspectives, and suggestions for improvement.

\subsection{Evaluation Results for Overall LUIDA Usage}

The following presents the results of our quantitative assessment and a summary of the researchers' interview responses.
A complete list of the actual responses can be found in Sec. 2 of the Supplementary Material, ``Interview Responses from VR Researchers Participating in the Usability Evaluation.''

\subsubsection{Quantitative Assessment}
Regarding the overall process of using LUIDA, the SUS score was 73.75 $\pm$ 13.39, interpreted as `Good' among `Worst Imaginable', `Awful', `Poor', `OK', `Good', `Excellent' and `Best Imaginable' \cite{sus_interpret}.

The NASA-TLX score was 24.11 $\pm$ 21.01, interpreted as `Medium' among `Low' (0--9), `Medium' (10--29), `Somewhat high' (30--49), `High' (50--79) and `Very high' (80--100) \cite{workload_interpret}.

\subsubsection{Key Benefits}
The interviews revealed the following key benefits of LUIDA, with participants' responses available in Sec. 2.1 of the Supplementary Material.
\begin{itemize}
    \item \textbf{Automation and Integration of Experimental Processes:} The automation of participant recruitment and data collection streamlines cumbersome tasks typically associated with traditional laboratory experiments.
    \item \textbf{Parallel Experiment Execution:} The system manages the execution of the experiment autonomously, without direct involvement of the researcher.
    \item \textbf{Enhanced Accessibility}: Remote participation allows access to diverse groups of participants, including those from rural and international locations.
    \item \textbf{Improved Information Management:} The Web Console's management of questionnaires and collected data improves efficiency and minimizes procedural errors.
\end{itemize}

\subsubsection{Challenges and Improvement Suggestions}
Participants identified challenges and potential improvements, with their actual responses available in Supplementary Material Sec. 2.2.

\begin{itemize}
    \item \textbf{Platform-Dependent User Base:} Dependency on Cluster limits the participant pool and may introduce bias. To address this, participants suggested improving support for users unfamiliar with Cluster to broaden the participant base and reduce platform dependency.
    \item \textbf{Challenges in Environmental Monitoring:} Verifying participant engagement and proper adherence to instructions proved difficult. To resolve this, participants suggested real-time support features, including automated alerts, Q\&A modules, and recording capabilities for post-experiment review.
\end{itemize}

\subsection{Evaluation Results for the Implementation Template}

\subsubsection{Quantitative Assessment}
The SUS score was 44.17 $\pm$ 11.80, which falls between `Poor' and `OK' \cite{sus_interpret}.
The NASA-TLX score was 63.33 $\pm$ 15.45, interpreted as `High' \cite{workload_interpret}.

\subsubsection{Strengths and Challenges}
Participants suggest that the main strength of the Implementation Template is the integration of functions for rational experimentation, with their actual responses available in Sec. 3.1 of the Supplementary Material.
\begin{itemize}
    \item \textbf{Integrated Approach:} The template eliminates the need to set up a dedicated server, allowing experiments to be run online directly on the platform.
    \item \textbf{Built-in Functionalities:} Automated questionnaire generation, state management, direct data recording, and CCK integration simplify experimental design and reduce manual overhead.
\end{itemize}

However, participants noted that there were challenges with complexity, debugging, and interface design:
\begin{itemize}
    \item \textbf{Complexity and Learning Curve:} Although powerful, the state management feature was considered overly complex for simpler experimental designs and debugging, especially for users unfamiliar with CCK or scripting.
    \item \textbf{User Interface (UI):} Cluttered layouts and lack of intuitive visual cues affected usability.
    \item \textbf{Partial No-Code Interface:} Frequent switching between coding and no-code user experiences during implementation potentially introduced confusion.
\end{itemize}

\subsubsection{Improvement Suggestions}
Participants suggested several solutions to address these challenges, with their actual responses available in Sec. 3.2 of the Supplementary Material.

\begin{itemize}
    \item \textbf{Modularization:} Restructure the system to import only the necessary components (questionnaire, data upload, state management) rather than working with a monolithic package.
    \item \textbf{Improving UI:} Adopt a node-based or table-driven GUI with color-coded elements for clearer visualization of state transitions.
    \item \textbf{Operation Streamlining:} Add copy-and-paste functionality for state settings and streamline controls to reduce excessive clicking (e.g., [P4]'s suggestion for a Gantt chart style layout).
    \item \textbf{Enhanced Tutorials and Documentation:} Improve documentation with video tutorials, sample scenes, and detailed integration guides to reduce barriers to entry and facilitate user learning.
\end{itemize}

\subsection{Discussion}
Researchers highly valued LUIDA's key features: automated recruitment, parallel execution, and streamlined data or questionnaire management, for their contributions to reducing workload, and increasing accessibility. The system was specifically praised for bridging the gap between urban and rural research institutions and removing barriers related to participants' nationality or cultural background.
While concerns regarding environmental control and platform-specific demographics were raised, the overall benefits outweighed these challenges, highlighting advantages over traditional laboratory experiments. This finding is supported by the high usability and moderate workload ratings for the overall LUIDA experience. To address these limitations, interview responses suggested expanding support to non-Cluster users and enhancing environmental control features.

In contrast, while the Implementation Template received positive feedback for its functionality, the quantitative evaluation indicated unsatisfactory usability and a high workload. Feedback from interviews indicated that these issues were primarily due to the complexity of the state management functionality and the cluttered interface. This highlighted significant improvement needs, with interview feedback suggesting solutions such as component modularization to reduce complexity and improve usability.

This user study also includes limitations in the type of experiments we asked the researchers to implement. For instance, the validation used pre-existing experimental designs, and future work should also validate the implementation of completely novel experimental systems. Furthermore, a future study could compare the effort and time required to implement the same experiment with and without using LUIDA to more directly quantify its benefits.

In summary, our findings show that LUIDA offers significant benefits through automated recruitment, parallel experiment execution, and efficient questionnaire and data management. These features increase accessibility and reduce barriers in VR research, addressing RQ1: "What benefits does this framework offer researchers in conducting experiments?". In addition, the successful implementation of all three experimental designs by researchers suggests that LUIDA can accommodate a wide variety of VR experiments (RQ2). Although challenges remain in environmental control, platform dependency, and complexity of Implementation Templates, they appear to be addressing them through future improvements.

\section{Large-Scale Validation with Replicated Experiments}\label{sec:user-study-online}
To evaluate how effectively LUIDA can promote experiment participation (RQ3) and the validity of LUIDA to host experiments (RQ4), we implemented and uploaded the three experiments described in Sec. \ref{sec:exp-to-replicate} using LUIDA.
These experiments were made publicly available for execution with participants outside the lab, and then we examined whether the data collected were consistent with the original experiments from which they were replicated.

\subsection{Experiment Procedure}
To recruit participants, we made LUIDA publicly available on the Cluster for one week, inviting users to the Recruitment World through Cluster's official events and social media. Participants consented to participation and were required to use VR devices and complete a questionnaire about their age, gender, and VR experience before participating in an experiment. The total number of participations during this fixed period served as a measure of how effectively LUIDA could promote participation.

After the one-week public period, we closed access to LUIDA and offered participants a 2,000 JPY (approximately 13 USD) Amazon gift card for each completed experiment (up to 6,000 JPY (approximately 40 USD) for participation in all three experiments). The results of each experiment were analyzed using the same analytical approaches used in the original studies to verify the degree of alignment.

\begin{table}[h]
    \centering
    \caption{VR devices used in replicated experiments. Oculus/Meta devices are grouped. Devices not explicitly noted as Standalone are PCVR devices used via SteamVR. }
    \label{tab:headset_types_all_exp}
    \begin{tabular}{lrrr}
        \hline
        \multirow{3}{*}{Device} & \multirow{3}{*}{\makecell[r]{Count in Hand\\Redirection\\Experiment}} & \multirow{3}{*}{\makecell[r]{Count in\\Proteus Effect\\Experiment}} & \multirow{3}{*}{\makecell[r]{Count in\\3D Fitts' Law\\Experiment}} \\\\\\ \hline
        \multirow{2}{*}{\makecell[l]{Meta Quest 2/3\\(Standalone)}} & \multirow{2}{*}{66} & \multirow{2}{*}{61} & \multirow{2}{*}{76} \\\\ \hline
        Meta Quest 2 & 64 & 63 & 66 \\ \hline
        Meta Quest 3 & 37 & 33 & 38 \\ \hline
        Meta Quest Pro & 2 & 2 & 2 \\ \hline
        Meta Quest 3S & 1 & 1 & 2 \\ \hline
        Oculus Rift S & 1 & 1 & 1 \\ \hline
        Oculus Rift CV1 & 1 & 1 & 1 \\ \hline
        VIVE Pro MV & 3 & 3 & 3 \\ \hline
        Vive XR Elite & 2 & 2 & 2 \\ \hline
        Vive & 1 & 1 & 1 \\ \hline
        Vive MV & 1 & 1 & 1 \\ \hline
        Pimax Crystal & 1 & 1 & -- \\ \hline
        PICO 4 & 2 & 2 & 1 \\ \hline
        Beyond & 1 & 1 & 1 \\ \hline
        Valve Index & 1 & 1 & 1 \\ \hline
        Unknown & 1 & -- & 1 \\ \hline
    \end{tabular}
\end{table}

\subsection{Hand Redirection Experiment}

\subsubsection{Measurements}
As detailed in Sec.~\ref{sec:exp-to-replicate-hand}, participants answered whether they perceived that the virtual hand moved faster or slower than their physical hand.
Each participant's answers were recorded and then fit to a psychometric sigmoid function to derive three key perceptual metrics:
\begin{itemize}
    \item \textbf{Faster Threshold:} The minimum gain at which the virtual hand movement is perceived as faster than the participant's actual hand.
    \item \textbf{Slower Threshold:} The maximum gain at which the virtual hand movement is perceived as slower than the participant's actual hand.
    \item \textbf{Point of Subjective Equality (PSE):} The gain level at which the virtual hand speed is perceived as being equal to the actual hand speed.
\end{itemize}

\subsubsection{Hypotheses}
We hypothesized that the Faster and Slower Thresholds measured in our experiment would be consistent with those reported in the original study.
Since the original study did not report standard deviations or confidence intervals,
we hypothesized that both the Faster (\textbf{HR-H1}) and Slower (\textbf{HR-H2}) Thresholds from our experiment would have confidence intervals containing the original study's values.

\subsubsection{Participants}
A total of 185 participants completed the hand redirection experiment. The VR devices they used are listed in Table~\ref{tab:headset_types_all_exp}.
After excluding 22 non-compliant participants (11.89\% of the whole dataset) who consistently selected only "faster" or "slower" in all trials, valid data were obtained from 163 participants (106 males, 45 females, 12 others; mean age = 32.012 $\pm$ 9.187).

\subsubsection{Results}
Table~\ref{tab:DT} lists the detection thresholds from both the original study and our replicated experiment.
Except for PSE, both the Faster and Slower Thresholds from the original study fell within the confidence intervals of our results.
This finding supports both \textbf{HR-H1} and \textbf{HR-H2}, suggesting that experiments conducted on LUIDA demonstrate sufficient validity in the field of perception research.

\setlength{\textfloatsep}{0pt}
\begin{table}
    \centering
    \caption{Detection threshold results in gain-based hand redirection}
    \label{tab:DT}
    \begin{tabular}{cccccc}
        \toprule
         & \makecell{Mean\\(Ours)} & \makecell{SD\\(Ours)} & \makecell{Confidence\\Interval (Ours)} & \makecell{Reported Value\\by Zenner \etal} \\
         \hline
        Faster & 1.12 & 0.596 & (1.032, 1.217) & 1.07 \\
        PSE & 1.035 & 0.334 & (0.983, 1.087) & 0.98 \\
        Slower & 0.945 & 0.686 & (0.839, 1.051) & 0.88 \\
        \bottomrule
    \end{tabular}
\end{table}

\subsection{Proteus Effect Experiment}

\subsubsection{Measurements}
During the drumming period (detailed in Sec.~\ref{sec:exp-to-replicate-proteus}), participants' body movement data were recorded each frame and analyzed using principal component analysis to calculate \textit{d95}, an indicator of ``performativity'' as defined in the original study.

Unlike the original study which tracked multiple body parts, Cluster by default only tracks the HMD and controllers (although it supports additional trackers such as HTC Vive trackers\footnote{\url{https://www.vive.com/accessory/tracker3/}}, we chose not to require them to lower the participation burden), limiting our analysis to head and hand positions.
We therefore conducted an additional analysis using \textit{total displacement} (sum of displacements for head and hands throughout the task) as a supplement.

\subsubsection{Hypotheses}
We hypothesized that our replicated experiment would yield results consistent with those of the original study.
Based on the original findings, we formulated parallel hypotheses for both \textit{d95} (PE-H1a, H2a, H3a) and \textit{total displacement} (PE-H1b, H2b, H3b):
\begin{itemize}
    \item \textbf{PE-H1a/b:} The metric for FL avatar participants does not differ significantly from baseline.
    \item \textbf{PE-H2a/b:} CD avatar participants exhibit higher values than baseline.
    \item \textbf{PE-H3a/b:} CD avatar participants exhibit higher values than FL avatar participants.
\end{itemize}

\subsubsection{Participants}
A total of 174 participants joined the Proteus Effect experiment, and the VR devices they used are listed in Table~\ref{tab:headset_types_all_exp}.
Participants were randomly assigned to one of two experimental groups: the Formal Light-skinned (FL) group or the Casual Dark-skinned (CD) group.
Each participant completed tasks first in a baseline condition and then in a condition where they embodied their assigned avatar, as described in Sec.~\ref{sec:exp-to-replicate-proteus}.
Among the participants, 16 dropped out (9.20\% of the whole dataset) during the transition from the baseline world (without avatar) to the avatar display world.
This resulted in valid data from 158 participants (102 males, 45 females, 11 others; mean age = 32.665 $\pm$ 9.331).

\subsubsection{Results}
A Shapiro-Wilk test indicated that the \textit{d95} data were not normally distributed ($p<0.05$).
We conducted Wilcoxon signed-rank tests to compare the baseline and embodied conditions within the CD and FL groups.
To compare the groups themselves, we applied Wilcoxon rank-sum tests to both their baseline and embodied conditions.

The results are presented in Table~\ref{tab:d95}.
In the FL group, there was no significant difference between the baseline and embodied conditions, supporting \textbf{PE-H1a}.
The CD group showed a significantly higher \textit{d95} in the embodied condition than in the baseline condition, supporting \textbf{PE-H2a}.
However, the absence of significant difference between the embodied conditions of the FL and CD groups fails to support \textbf{PE-H3a}.

Similarly, the Shapiro-Wilk test showed that the \textit{total displacement} metric was not normally distributed ($p<0.05$), so we applied the same set of non-parametric analyses.
As shown in Table~\ref{tab:total-displacement}, the results mirrored the \textit{d95} findings.
The FL group showed no significant difference between the baseline and embodied conditions (\textbf{PE-H1b} supported), while the CD group exhibited significantly higher \textit{total displacement} in the embodied condition than in the baseline condition (\textbf{PE-H2b} supported).
Again, the comparison between the embodied conditions of the FL and CD groups revealed no significant difference, failing to support \textbf{PE-H3b}.

In summary, the analyses of both \textit{d95} and \textit{total displacement} yielded consistent results.
These findings suggest that our experiment on the LUIDA platform partially replicated the Proteus effect observed by Kilteni et al.~\cite{drumming_proteus}, indicating LUIDA's potential to maintain experimental validity in avatar embodiment and related cognitive science contexts.

\begin{table}
    \centering
    \caption{Analysis results for \textit{d95} in the baseline and avatar conditions of the CD and FL groups}
    \label{tab:d95}
    \begin{tabular}{cccccc}
        \toprule
        Group & base & exp & \textit{p} & Cohen's \textit{r} & n \\
        \hline
        CD & 6.11 $\pm$ 0.968 & 6.42 $\pm$ 0.952 & .011 & 0.202 & 80 \\
        FL & 6.19 $\pm$ 0.869 & 6.36 $\pm$ 0.852 & .202 & 0.102 & 78 \\
        \textit{p} & .505 & .475 &  & \\
        Cohen's \textit{r} & 0.0530 & 0.0568 &  & \\
        \bottomrule
    \end{tabular}
\end{table}

\begin{table}
    \centering
    \caption{Analysis results for total displacement in the baseline and avatar conditions of the CD and FL groups (r: Cohen's r)}
    \label{tab:total-displacement}
    \begin{tabular}{@{}cccccc@{}}
        \toprule
        Group & base & exp & \textit{p} & \textit{r} & n \\
        \hline
        CD & 320.721 $\pm$ 135.180 & 351.078 $\pm$ 141.451 & .004 & 0.231 & 80 \\
        FL & 326.903 $\pm$ 127.956 & 321.138 $\pm$ 141.188 & .158 & 0.113 & 78 \\
        \textit{p} & .130 & .122 &  & \\
        \textit{r} & 0.121 & 0.123 &  & \\
        \bottomrule
    \end{tabular}
\end{table}

\subsection{3D Fitts' Law Experiment}

\subsubsection{Measurements}
The primary dependent variable was Movement Time (MT). Independent variables included the target's size and its 3D position, from which spatial variables (distance, depth, inclination angle, azimuth angle) and Index of Difficulty (ID) were derived.
Additionally, we collected demographic data (age, gender), prior VR experience, and subjective ratings of presence via questionnaires.

\subsubsection{Hypotheses}
We aimed to validate whether our platform has experimental validity equivalent to previous studies in this field.
Specifically, we intended to confirm that data obtained from our platform could be predicted by existing models, including the model proposed by Clark \etal, as well as other models introduced and compared in their study (Hoffmann, Murata \& Iwase, Cha \& Myung, Machuca \& Stuerzlinger), at a level equal to or greater than that reported by Clark \etal
We then formulated three hypotheses:
movement time is significantly predicted by the same or more variables (\textbf{FL-H1}); the model fit, reflected in adjusted coefficients of determination (Adj.~$R^2$), is equivalent or higher (\textbf{FL-H2}); and the standard error of the regression ($SE$) is equivalent or lower (\textbf{FL-H3}).

\subsubsection{Participants}
A total of 197 participants (131 males, 54 females, 12 others; mean age = 32.352 $\pm$ 9.423) completed the experiment. None of the participants dropped out or were noncompliant with the task instructions. The VR devices they used are listed in Table~\ref{tab:headset_types_all_exp}.

\subsubsection{Results}
Following Clark~\etal~\cite{fitts_3d}, we excluded trials with movement times greater than 1.5 IQR above the upper quartile as outliers.
Linear regression analyses were then performed, guided by Saffo~\etal~\cite{remote_exp_social_vr}, to model movement time using variables from the existing models listed in Clark~\etal’s study (see Sec.~\ref{sec:3d-fitts-law-intro}).

The results in Table~\ref{tab:fitts-results} show that all spatial variables in each model, including Index of Difficulty, depth, inclination angle, azimuth angle, and target size $\times$ inclination angle, significantly predicted movement time, supporting \textbf{FL-H1}.
Meanwhile, all models explained a moderate portion of the variance, but Adj. $R^2$ ($37.3\sim38.8\%$) was lower than Clark \etal's work ($50.1\sim64.5\%$), failing to support \textbf{FL-H2}.
In contrast, SE ($285.779\sim289.336~ms$) was lower than Clark \etal's work ($364.162\sim431.963~ms$), supporting \textbf{FL-H3}.

In addition, Saffo~\etal also conducted a similar replication of Clark~\etal on VRChat but with experimenter's supervision and participants embodying an avatar during the experiment.
A post-hoc comparison with Saffo~\etal's work shows that our work identified more significant variables but a worse model fit than theirs (lower Adj. $R^2$ and higher SE; see Table~\ref{tab:fitts-results}).

\begin{table*}[ht]
\centering
\caption{Models evaluated by Clark~\etal~\cite{fitts_3d} and our replication ($ID$: Index of Difficulty, $\theta$: inclination angle, $\varphi$: azimuth angle).}
\label{tab:fitts-results}
\resizebox{\textwidth}{!}{%
\begin{tabular}{l l r r r r r r r r r r r r r}
\toprule
\multirow{2}{*}{Model} & \multirow{2}{*}{Parameters} & \multicolumn{2}{c}{Estimate} & \multicolumn{2}{c}{SE} & \multicolumn{2}{c}{sr} & \multicolumn{2}{c}{$p$} & \multicolumn{2}{c}{$R^2$} & \multicolumn{2}{c}{SE (ms)} & \multicolumn{1}{c}{Adj. $R^2$} \\
\cmidrule(r){3-4} \cmidrule(r){5-6} \cmidrule(r){7-8} \cmidrule(r){9-10} \cmidrule(r){11-12} \cmidrule(r){13-14}\cmidrule(r){15-15}
 & & Clark & Ours & Clark & Ours & Clark & Ours & Clark & Ours & Clark & Ours & Clark & Ours & Ours \\
\midrule
\multirow{2}{*}{\makecell[l]{Hoffmann\\(Fitts') \cite{fitts_original}}} & Intercept & 804.287 & 481.327 & 123.000 & 4.633 & - & - & - & - & 0.501 & 0.373 & 427.450 & 289.336 & 0.373 \\
 & ID & 411.940 & 130.121 & 42.440 & 1.243 & 0.708 & 0.611 & $<$.001 & $<$.001 & - & - & - & - & - \\
\midrule
\multirow{3}{*}{\makecell[l]{Murata \&\\Iwase \cite{fitts_murata_iwase}}} & Intercept & 805.404 & 480.710 & 123.907 & 4.639 & - & - & - & - & 0.501 & 0.373 & 429.697 & 289.295 & 0.373 \\
 & ID & 411.760 & 130.101 & 42.678 & 1.243 & 0.707 & 0.611 & $<$.001 & $<$.001 & - & - & - & - & - \\
 & $\sin(\varphi)$ & 8.506 & 7.835 & 61.130 & 3.142 & 0.010 & 0.015 & .890 & .013 & - & - & - & - & - \\
\midrule
\multirow{4}{*}{\makecell[l]{Cha \&\\Myung \cite{fitts_cha_myung}}} & Intercept & 819.971 & 480.251 & 152.789 & 4.587 & - & - & - & - & 0.501 & 0.387 & 431.963 & 286.036 & 0.387 \\
 & ID & 411.846 & 130.223 & 42.060 & 1.229 & 0.707 & 0.611 & $<$.001 & $<$.001 & - & - & - & - & - \\
 & $\sin(\varphi)$ & 6.429 & 7.413 & 62.735 & 3.107 & 0.008 & 0.014 & .919 & .017 & - & - & - & - & - \\
 & $\theta$ & -0.290 & 1.026 & 1.759 & 0.050 & -0.012 & 0.119 & .135 & $<$.001 & - & - & - & - & - \\
\midrule
\multirow{3}{*}{\makecell[l]{Machuca \&\\Stuerzlinger\\\cite{fitts_machuca}}} & Intercept & 795.413 & 480.871 & 118.073 & 4.582 & - & - & - & - & 0.545 & 0.387 & 410.202 & 286.100 & 0.387 \\
 & ID & 415.547 & 130.224 & 40.742 & 1.230 & 0.713 & 0.611 & $<$.001 & $<$.001 & - & - & - & - & - \\
 & Depth & 3.505 & 1.215 & 1.164 & 0.059 & 0.211 & 0.118 & .003 & $<$.001 & - & - & - & - & - \\
\midrule
\multirow{4}{*}{\makecell[l]{Clark~\etal\\\cite{fitts_3d}}} & Intercept & 732.369 & 481.012 & 105.456 & 4.577 & - & - & - & - & 0.645 & 0.388 & 364.162 & 285.779 & 0.388 \\
 & ID & 433.127 & 130.209 & 36.319 & 1.228 & 0.741 & 0.611 & $<$.001 & $<$.001 & - & - & - & - & - \\
 & $\theta$ & -2.681 & 1.352 & 1.170 & 0.072 & -0.142 & 0.119 & .024 & $<$.001 & - & - & - & - & - \\
 & Size$~\times~\theta$ & 0.336 & -0.017 & 0.060 & 0.003 & 0.346 & -0.036 & $<$.001 & $<$.001 & - & - & - & - & - \\
\bottomrule
\end{tabular}%
}
\end{table*}

\subsubsection{Discussion}

The support for \textbf{FL-H1} and \textbf{FL-H3} suggests our platform collects high-fidelity data; the lower $SE$ implies less noise in the measurements, and the larger number of significant variables indicates a higher sensitivity to subtle behavioral effects.
However, the lower Adj. $R^2$ value suggests that the established models explain less of the variance within our dataset.
Key factors discussed below likely explain these differences in model fit across ours and the prior works.

\paragraph{Hardware Specifications} The original study by Clark~\etal utilized the Oculus Rift CV1, and the replication by Saffo~\etal employed hardware from a similar generation (e.g., HTC Vive, Valve Index, Oculus Rift S).
Our study, however, was conducted on more modern headsets (predominantly Meta Quest 2 and 3) that feature significantly higher resolutions and different display technologies (LCD versus OLED).
This technological disparity likely introduced new sources of variance in user performance that are not captured by the original models.
The relatively farther hardware specifications in our study plausibly explains the weaker model fit to our data compared to both Clark~\etal and Saffo~\etal

\paragraph{Participant Factors} Our large and diverse participant pool ($N=197$), who participated remotely using their personal equipment, inherently introduces more variability than is typical in a controlled laboratory experiment (Clark~\etal) or where experimenter is supervising (Saffo~\etal).

\paragraph{Use of Avatars} Additionally, Saffo~\etal suggested that their use of embodied avatars may have enhanced depth cues, a factor which could have contributed to less noise and thereby a lower SE in their work compared to ours.

In summary, the comparison with the original study demonstrates LUIDA experimental validity by replicating core findings with high precision, evidenced by the low standard error and the detection of additional subtle variables.
The lower model fit is not interpreted as a failure of the platform but rather as a significant finding for modern VR research: models developed on older, homogenous hardware do not fully account for the performance variance introduced by the diversity of today's user base and technology.
This result underscores our system's capacity for capturing nuanced user performance data in more ecologically valid, ``in-the-wild'' settings.

A post-hoc comparison with a similar replication study implies (1) the use of avatars as spatial cues, which is not a fundamental feature of LUIDA, and (2) enhanced participant control.
This need for greater control offers critical suggestions for LUIDA's future improvement or proper usage.
For instance, a supervision feature could be introduced.
To manage device heterogeneity, researchers could either restrict device types before participation or group data by device during post-experiment analysis to achieve higher control.

\subsection{Discussion}
\subsubsection{Experimental Validity with Low Data Loss across Domains}
All replicated experiments generated results closely matching the original studies despite participants using diverse VR hardware:
the hand redirection experiment yielded consistent DTs (\textbf{HR-H1}, \textbf{HR-H2}), the Proteus effect experiment partially replicated original findings (\textbf{PE-H1a/b}, \textbf{PE-H2a/b}), and the 3D Fitts' Law experiment aligned with existing models with less standard errors (\textbf{FL-H1}, \textbf{FL-H3}).
Given that the original studies are widely accepted and considered valid, the consistency between our results and the originals suggests that LUIDA maintains adequate validity without introducing uncontrolled variables that could impact outcomes.

Furthermore, traditional crowdsourcing approaches often result in a high proportion of data having insufficient quality.
For instance, Brühlmann~\etal's experiment reported that 45.9\% of the participants exhibited careless behavior during the experiment~\cite{BRUHLMANN2020100022}.
Similarly, Pickering~\etal reviewed previous studies that conducted crowdsourcing experiments, finding that some studies had to discard 16\%--49\% of the collected data due to invalid or incomplete responses~\cite{Pickering02092021}.
In contrast, LUIDA achieved much lower data exclusion rates of 11.89\%, 9.20\%, 0\% across the three replicated experiments, respectively.
Such low proportions of invalid data further support the validity of LUIDA as an experimental framework.
We suggest that this outcome is partly because participants were required to equip VR devices to join the experiments, which formed a barrier to entry and filtered out non-serious or inattentive users.
Therefore, we argue that even if the number of metaverse users increases in the future, the data quality on LUIDA is unlikely to be severely impacted, indicating LUIDA's sustainability.

Overall, we conclude that LUIDA sufficiently ensures experimental validity, effectively answering RQ4: ``Can this framework host experiments without compromising their validity compared to traditional laboratory experiments?''.
Besides, such validity was demonstrated across various research fields and devices, providing additional support for RQ2: ``Can this framework accommodate a wide variety of VR experiments?''.

\subsubsection{High Efficiency and Accessibility}
Despite aids by promotion, the ability to collect over 500 sets of data within a single week highlights how the platform’s automation and scalability significantly improve the efficiency of participant recruitment, experiment execution, and data collection.
This addresses RQ1: ``What benefits does this framework offer researchers in conducting experiments''.
The high number of participants also underscores LUIDA's high accessibility, which successfully encouraged participation in the experiments, answering RQ3: ``Can this framework encourage participant participation in experiments?''.

Taken together, the replication experiments demonstrated that LUIDA can reproduce key experimental effects observed in traditional in-lab studies, while also offering added benefits of automation, scalability, and accessibility.
This positions LUIDA as a robust alternative to conventional, lab-based methods for VR research.

\subsubsection{Challenges in Participant Dropout and Misunderstanding}
22 of 185 participants in the hand redirection experiment consistently selected either 'faster' or 'slower' in all trials, while 16 of 174 participants in the Proteus effect experiment dropped out during the world transition.
These data losses, while lower than those in traditional crowdsourcing approaches, likely resulted from misunderstandings of the experimental procedure despite the provided instructions.
For example, in the Proteus effect experiment, participants likely misinterpreted the appearance of a portal as the end of the study.
While LUIDA's continuous tracking allowed us to identify and exclude these cases to preserve validity, these losses underscore the persistent challenges of unsupervised remote research.
Future refinements could include system-level restrictions or enhanced automation to further mitigate it.

\subsubsection{Network Latency and Timing Precision}
As a networked platform, LUIDA introduces potential timing variability due to network latency.
While this did not impact our replicated experiments, studies requiring millisecond-precise timing should consider this limitation.
To address it, LUIDA's timestamped data logs (Sec.~\ref{subsubsec:implementation-constraints-all-platforms}, implemented in Sec.~\ref{par:data-integrity-privacy}) enable researchers to identify and filter network delayed data.
Future work could further mitigate latency by delegating the execution of certain processes to users' local devices.

\subsubsection{Benefits and Challenges of Metaverse Community Engagement}
The metaverse user community played an unexpected role in promoting experiments:
participants shared selfies from LUIDA's Recruitment World on social media, or introduced LUIDA to their friends within Cluster.
Many highlighted the ease of participation, supporting RQ3 from another aspect.
Moreover, this community-driven promotion may have encouraged participants to engage more actively in the experiments, as the information was shared by people they know.
This likely contributed to the low proportion of invalid or incomplete data, further reinforcing LUIDA's validity.

However, such casual sharing can also risk revealing experimental conditions, potentially challenging the validity of the experiments; yet, no such disclosures were identified during the week our experiments were public.
We recommend leveraging community-driven promotion while providing guidance to prevent disclosure of sensitive experimental details.

\section{Discussion}
\label{sec:discussion}
\subsection{Overall Benefits of LUIDA}
Our findings confirm LUIDA as a unified, structured and valid methodology for metaverse-based research, as well as an efficient alternative to traditional lab work.
By more effectively leveraging the inherent functionalities of the metaverse than prior attempts, LUIDA automates and parallelizes both participant recruitment and experiment execution.
These core benefits address critical challenges in VR studies, such as scalability, participant diversity, and the need for technical expertise or efforts, without compromising validity.
While its ability to remotely recruit large numbers of participants is comparable to traditional crowdsourcing, LUIDA demonstrates superior performance through lower data exclusion rates and better access to the target population of actual metaverse users, yielding findings that more accurately reflect the true ecosystem.

\subsection{Compensation for Data Loss and Experimental Control}

Our evaluation also identified areas for improvement in usability and flexibility to address more experimental needs.
One key challenge lies in data collection, particularly data loss caused by the lack of participant monitoring or support, which risks compromising experimental validity.
To address this, we plan to introduce system-level restrictions to mitigate common misunderstandings of experimental procedures (e.g., direct redirection instead of using world portals during world transitions).
It would also be beneficial to leverage methods from traditional crowdsourcing studies, such as developing a VR-adapted Instructional Manipulation Check~\cite{IMC} to detect noncompliance.
Additionally, based on user feedback, we aim to implement features that help researchers support participants, such as in-VR Q\&A forms, recording functionality, or allowing researchers to join ongoing sessions for monitoring.

Another risk is that if participants log off the platform, either deliberately or accidentally due to network problems, LUIDA currently does not natively support either automatic detection of lost data or session recovery upon their return, resulting in the loss of experimental state.
As a temporary alternative, researchers can manually verify whether a participant completed the session by checking the data log, as all collected data is timestamped.
To fundamentally address this issue, while metaverse platforms like Cluster provide APIs for saving and loading world states, which could be used by researchers within LUIDA's Implementation Template's customizable code blocks, a native, user-friendly data recovery and session resume feature should be included in future development.
This would allow researchers to more easily manage incomplete sessions without requiring custom scripting.

There are other aspects to consider in order to improve experimental control.
First, to prevent overly recruiting participants, experiments should automatically become invisible in the Recruitment World once participant quotas are reached.
Second, to ensure balanced sampling, the Experiment World could selectively reject participation based on self-reported demographics or detected hardware specifications.  
Finally, an automated testing feature would allow researchers to verify system functionality and compatibility within LUIDA.

\subsection{Refactoring for a More Lucid Implementation Template}

The quantitative evaluation exhibited limitations in usability and workload for the Implementation Template, with interview feedback revealing related issues. 
First, its cluttered user interface has negatively impacted usability and increased workload.
To address this, we plan a complete redesign using a node-based or table-driven layout for easier navigation.
Another challenge lies in its monolithic structure, which has posed difficulties for unfamiliar users struggling with multiple components at once.
To reduce hurdles due to this structure, we will refactor the template into modular but interoperable components, allowing researchers to adopt only the features they need.
These planned improvements are expected to enhance usability and reduce workload, supported by video tutorials and sample scenes to facilitate quicker onboarding.

\subsection{Leveraging Metaverse User Community}

Regarding the benefits and risks brought by the user community within the metaverse, one approach is to place a designated photo spot in the Recruitment World to encourage users to take and share photos, better leveraging community-driven promotion.
At the same time, clear warning signs should be displayed to specify which types of information can or cannot be disclosed, helping to prevent unintentional threats to experimental validity.

\section{Conclusion}
\label{sec:conclusion}

This paper introduced LUIDA, a novel framework as a structured methodology for VR/HCI experiments by leveraging advantages of commercial metaverse platforms.
Its core design unifies participant recruitment, execution, and data collection into a single, streamlined workflow.

A prototype developed on the metaverse platform Cluster was validated through two studies.
A usability evaluation with VR researchers confirmed LUIDA's ability to streamline experiment setup across multiple domains, while three large-scale replicated experiments demonstrated its capacity to rapidly recruit hundreds of participants and reproduce findings from original lab studies.
These results establish that LUIDA preserves experimental validity while offering transformative advantages in scale and automation for VR research.
We are continuing refinements to address data recovery, supervision, and Implementation Template usability, while also planning to incorporate ``Third Space'' design elements~\cite{third_place} into the Recruitment World to foster active communities and bridge research with society.

\section*{Conflict of Interest Statement}

The authors declare that the research was conducted in the absence of any commercial or financial relationships that could be construed as a potential conflict of interest.

\section*{Author Contributions}

YHH: Conceptualization, Formal analysis, Investigation, Methodology, Project administration, Software, Visualization, Writing – original draft, Writing – review \& editing.
SY: Methodology, Software, Visualization, Writing – original draft.
YHa: Conceptualization, Supervision, Validation, Writing – review \& editing.
YHi: Validation, Writing – original draft, Writing – review \& editing.
TN: Conceptualization, Funding acquisition, Supervision, Validation, Writing – review \& editing.
TH: Conceptualization, Funding acquisition, Project administration, Resources, Supervision, Validation, Writing – review \& editing.

\section*{Funding}
\noindent This work was supported by Moonshot Research \& Development Program (JPMJMS2013).


\section*{Data Availability Statement}
We have released LUIDA's current implementation by making the Recruitment World public on Cluster\footnote{\url{https://cluster.mu/w/29ac4e39-e7af-4467-b3ba-dea011473eb0}}, launching the Web Console\footnote{\url{https://luida.cluster.mu}} for researchers, and publishing the Implementation Template on GitHub\footnote{\url{https://github.com/cluster-lab/project-luida-bar}}.

\bibliographystyle{Frontiers-Harvard}
\bibliography{references}


\end{document}


\onecolumn
\firstpage{1}

\title[Supplementary Material]{{\helveticaitalic{Supplementary Material}}}

\maketitle

\section{Interface of the Web Console implemented in the LUIDA prototype}

\begin{figure}[H]
\centering
    \includegraphics[width=0.9\textwidth]{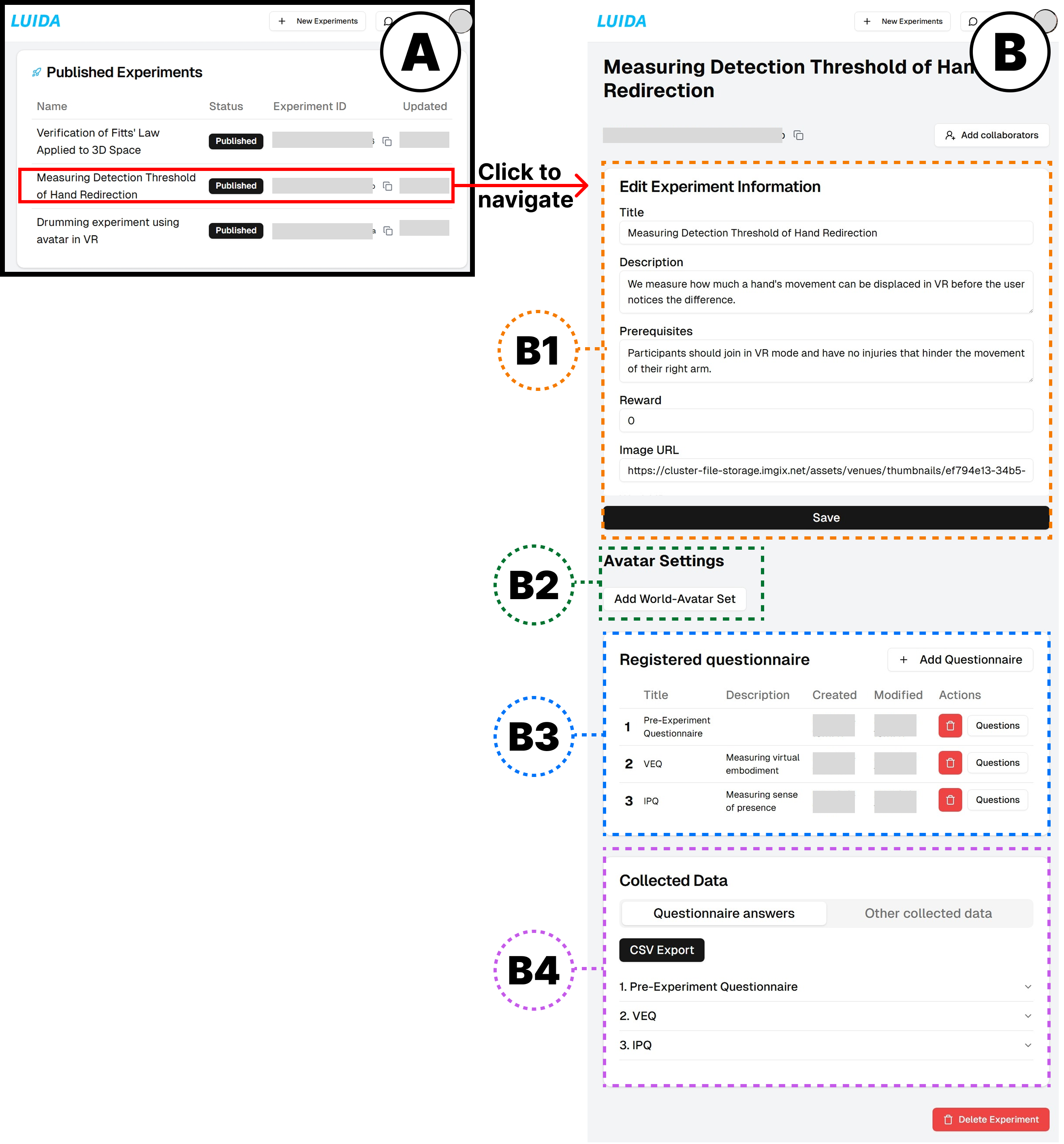}
    \caption{(A) Home page displaying a list of registered experiments; (B) Experiment creation/editing page containing: (B1) a form to edit experiment information, (B2) the avatar settings section  (B3) the questionnaires section, and (B4) the collected data section.}
\label{fig:console-exp-page}
\end{figure}

\begin{figure}[H]
\centering
    \includegraphics[width=\textwidth]{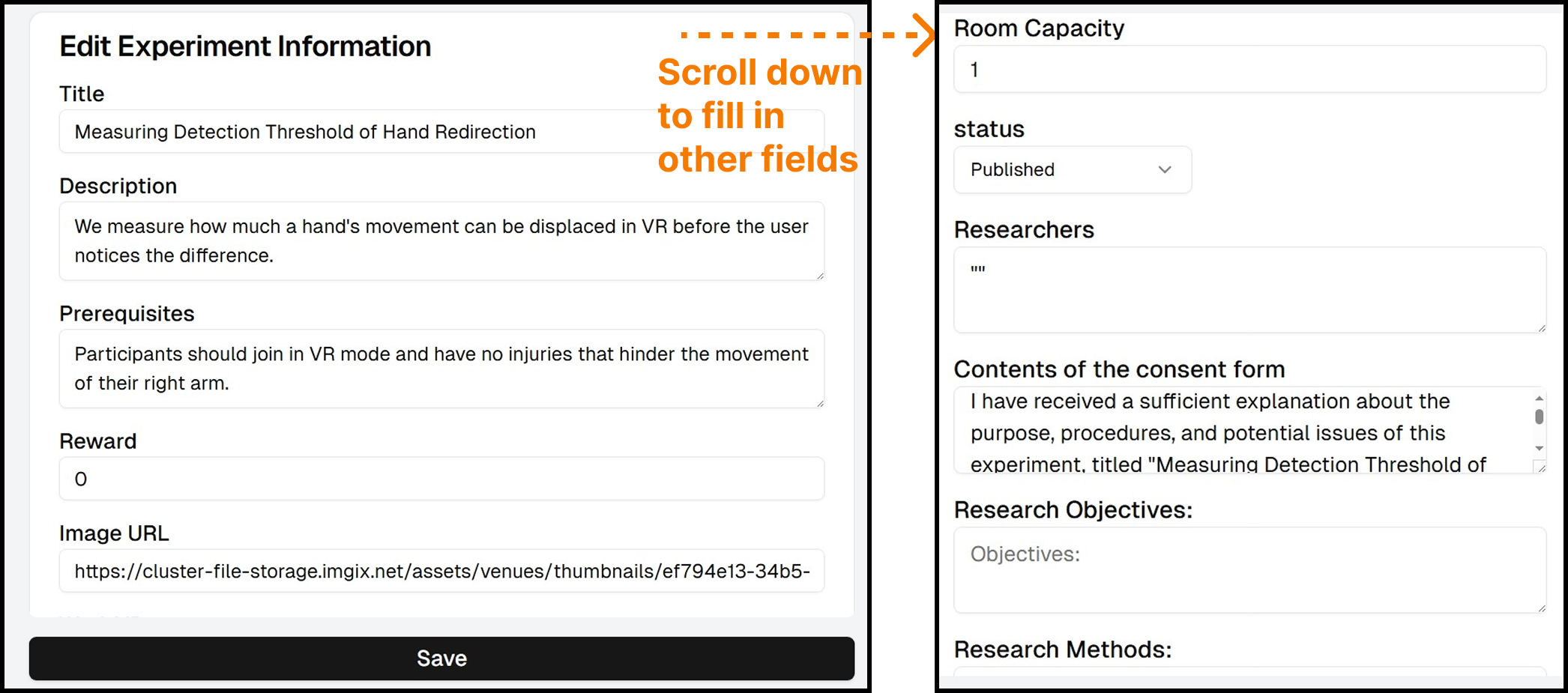}
    \caption{A scrollable form to edit experiment information, including title, description, prerequisites, reward, room capacity, contents of the consent form, etc.}
\label{fig:console-exp-detail-form}
\end{figure}

\begin{figure}[H]
\centering
    \includegraphics[width=0.7\textwidth]{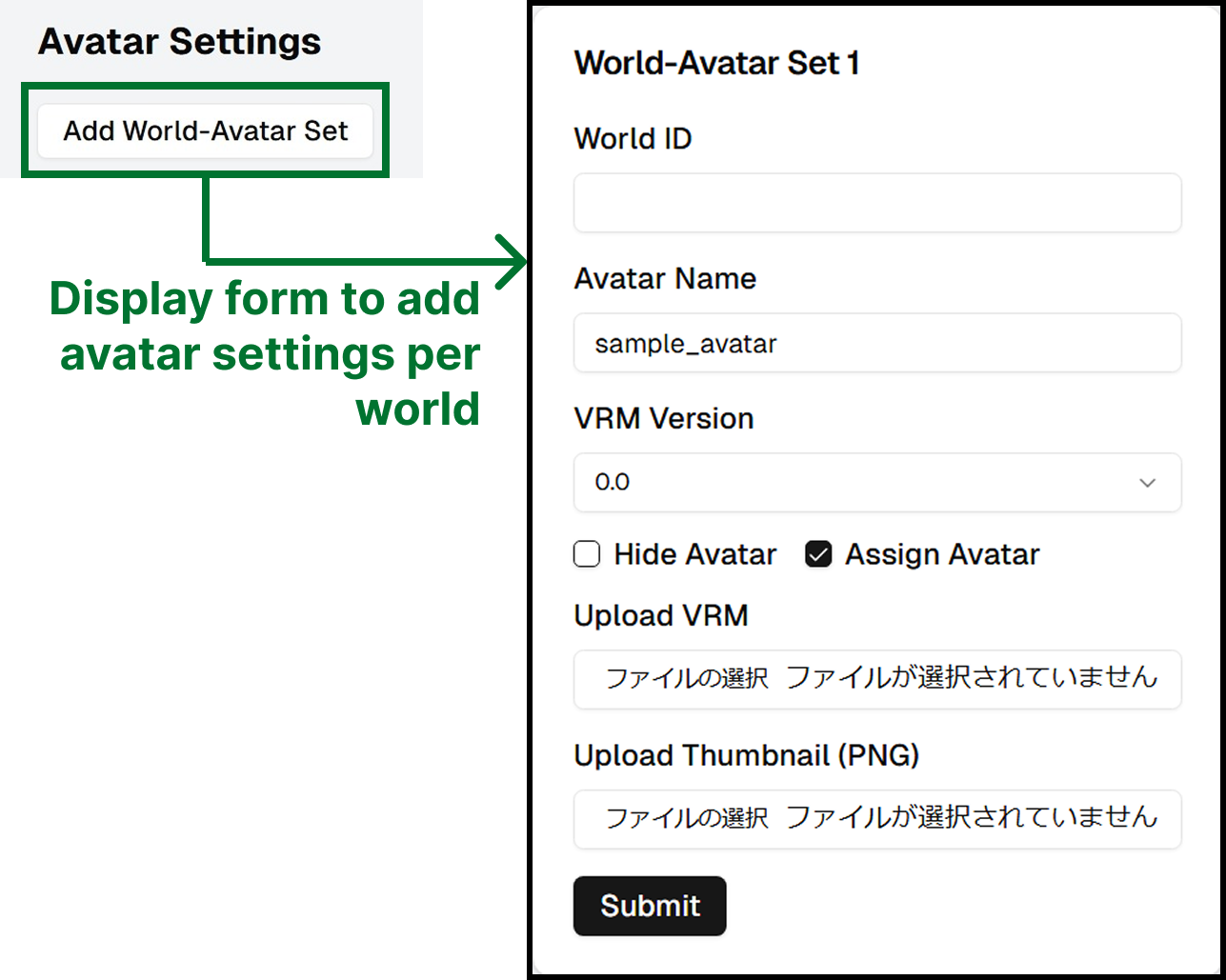}
    \caption{Avatar settings section with a button to add world–avatar sets. When clicked, the button opens a form that allows researchers to configure the avatar for each world, either by assigning an uploaded avatar (VRM file) or by hiding avatars (using a transparent avatar).}
\label{fig:console-avatar-settings}
\end{figure}

\begin{figure}[H]
\centering
    \includegraphics[width=\textwidth]{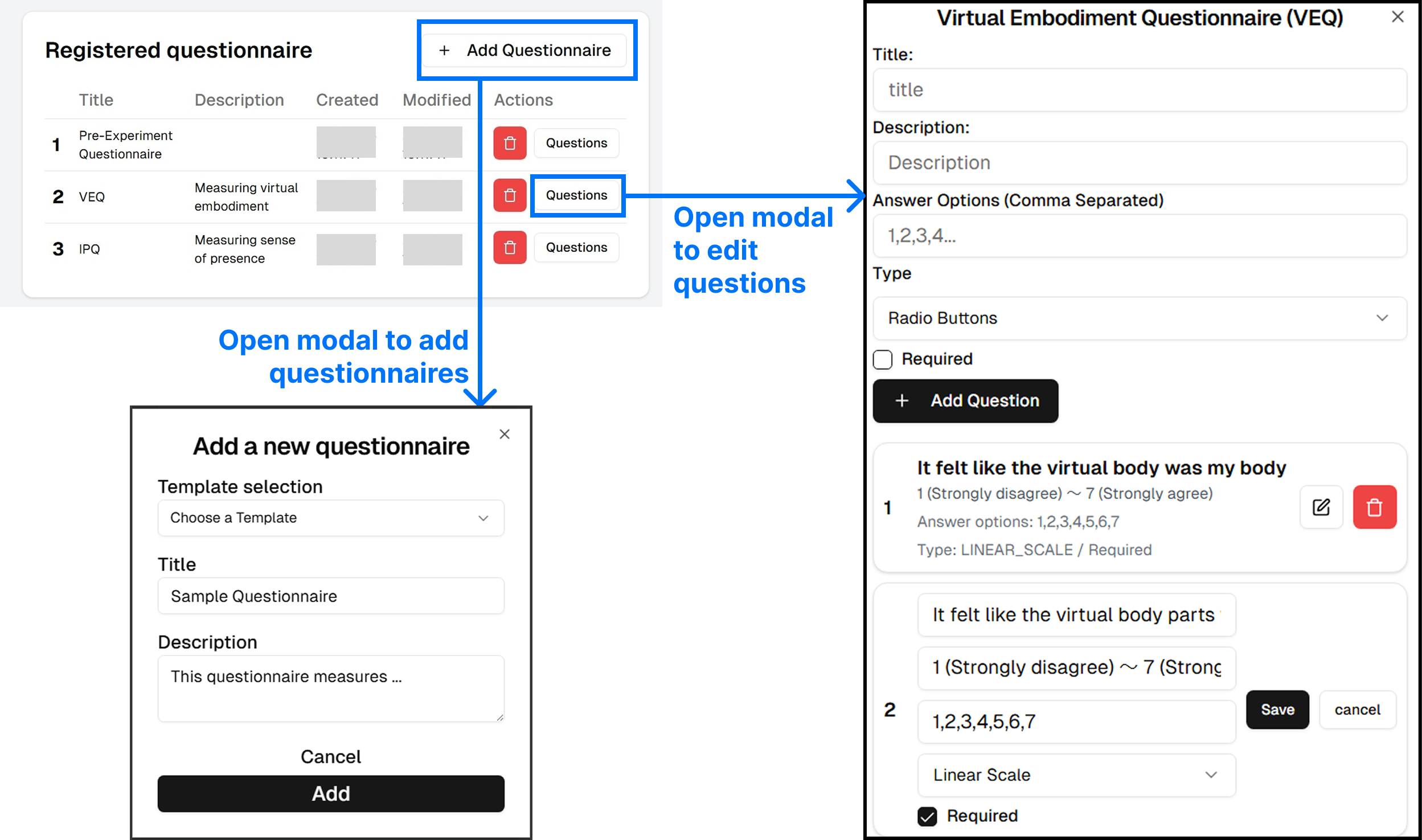}
    \caption{Questionnaires section with a button to add new questionnaires and a list of those added. Researchers can edit the question items within each questionnaire.}
\label{fig:console-questionnaires}
\end{figure}

\begin{figure}[H]
\centering
    \includegraphics[width=\textwidth]{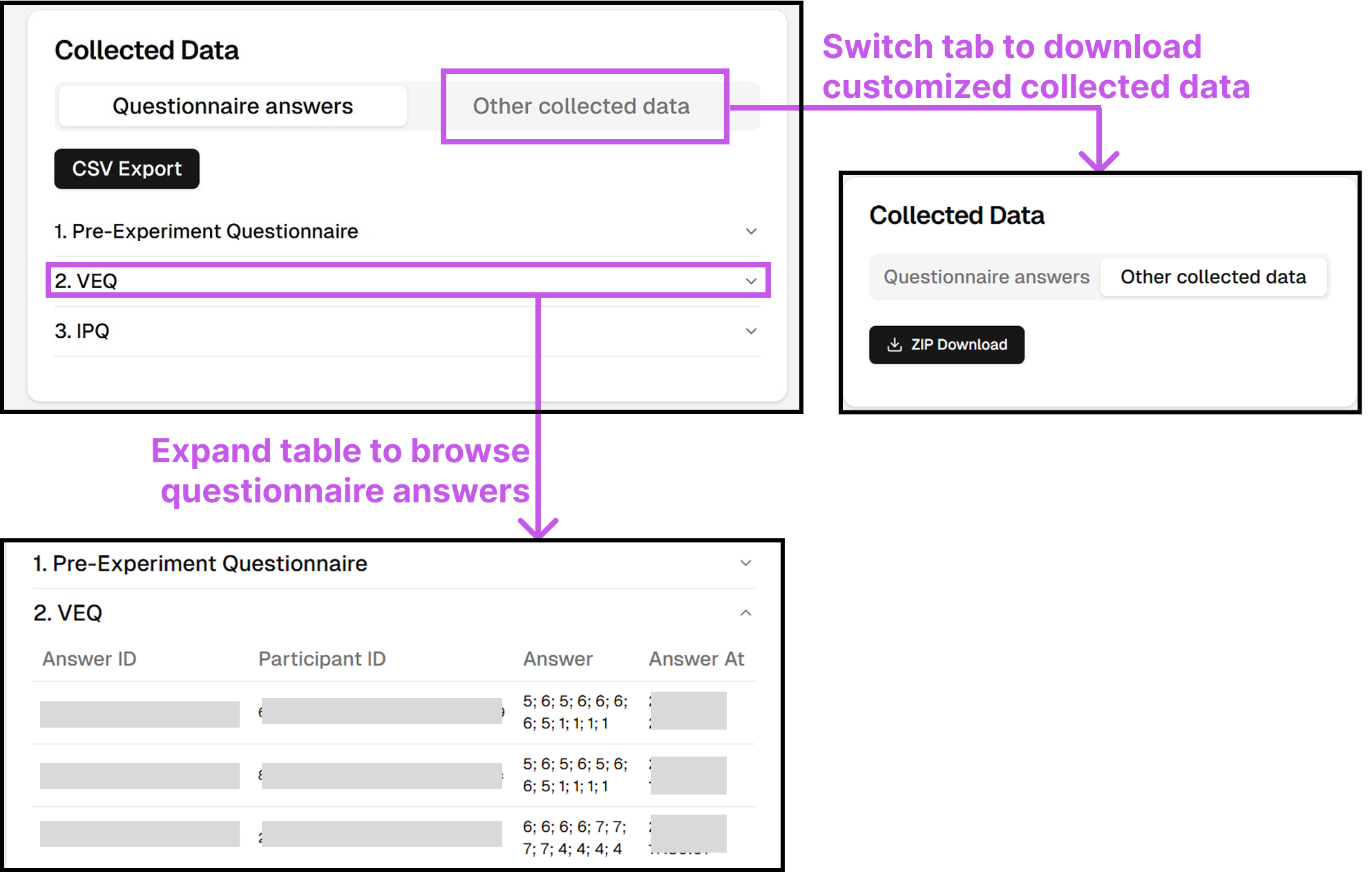}
    \caption{Collected data section with lists of responses for each questionnaire and a download link for other collected data.}
\label{fig:console-collected-data}
\end{figure}


\section{Interview Responses from VR Researchers Participating in the Usability Evaluation: Evaluation Results for Overall LUIDA Usage}

\subsection{Key Benefits}
\begin{itemize}
    \item \textbf{Automation and Integration of Experimental Processes:}
    \begin{itemize}
        \item \textit{"The ability to automatically gather participants online without recruiting them manually seems convenient. Recruiting participants so far had been such a hassle for me.
        " [P1]}
    \end{itemize}
    \item \textbf{Parallel Experiment Execution:}
    \begin{itemize}
        \item \textit{"It's convenient that experiments can run automatically without a researcher present." [P6]}
    \end{itemize}
    \item \textbf{Enhanced Accessibility}:
    \begin{itemize}
        \item \textit{"It allows access to people in rural areas and even from all over the world." [P2]}
        \item \textit{"Being able to recruit participants and conduct the experiment remotely, without requiring any actions from the researcher, makes things much easier." [P4]}
    \end{itemize}
    \item \textbf{Improved Information Management:}
    \begin{itemize}
        \item \textit{"No more risk of forgetting to collect questionnaires!" [P3]}
    \end{itemize}
\end{itemize}

\subsection{Challenges and Improvement Suggestions}
\begin{itemize}
    \item \textbf{Platform-Dependent User Base:}
    \begin{itemize}
        \item \textit{"Since participants are limited to Cluster users, there might be some bias." [P5]}
        \item \textit{"If a non-Cluster user has to use this system, they will need support." [P4]}
    \end{itemize}
    \item \textbf{Challenges in Environmental Monitoring:}
    \begin{itemize}
        \item \textit{"It's hard to confirm whether participants are actually doing the task correctly." [P1]}
        \item \textit{"There's no way to observe the participants, so it's difficult to check if they are doing the task properly." [P4]}
        \item \textit{"I would like a recording feature." [P6]}
    \end{itemize}
\end{itemize}

\section{Interview Responses from VR Researchers Participating in the Usability Evaluation: Evaluation Results for the Implementation Template}

\subsection{Strengths and Challenges}

\begin{itemize}
    \item \textbf{Integrated Approach:} 
    \begin{itemize}
        \item \textit{"It is convenient to be able to conduct online experiments without having to set up a server myself or implement network communication myself." [P1]}
        \item "\textit{It is nice to be able to complete questionnaire responses in the virtual environment without having to remove the HMD. I used to ask participants to answer questionnaires through Google Form on a tablet, which required them to repeatedly take off and put on the HMD. This made the experiment process too complicated." [P4]}
    \end{itemize}
    \item \textbf{Built-in Functionalities:}
    \begin{itemize}
        \item \textit{"The questionnaire generation feature is particularly helpful." [P5]}
        \item \textit{"CCK helps me facilitate settings for networking, VR locomotion, and interactions with objects." [P1]}
    \end{itemize}
\end{itemize}

\begin{itemize}
    \item \textbf{Complexity and Learning Curve:}
    \begin{itemize}
        \item \textit{"It is not friendly enough for users with no experience in state management." [P4]}
        \item \textit{"I'm still not familiar with ClusterScript and CCK... I can't understand the error messages" [P3]}
    \end{itemize}
    \item \textbf{User Interface (UI):}
    \begin{itemize}
        \item \textit{"Having all the features come together is not ideal... the UI is hard to read." [P1]}
        \item \textit{"Too many settings to configure, and it's unclear where and how to make changes." [P5]}
    \end{itemize}
\end{itemize}

\subsection{Improvement Suggestions}

\begin{itemize}
    \item \textbf{Modularization:} Restructure the system to import only the necessary components (questionnaire, data upload, state management) rather than working with a monolithic package [P1].
    \item \textbf{Improving UI:}
    \begin{itemize}
        \item \textit{"I would prefer color coding in the UI." [P1]}
    \end{itemize}
    \item \textbf{Operation Streamlining:} Add copy-and-paste functionality for state settings and streamline controls to reduce excessive clicking (e.g., [P4]'s suggestion for a Gantt chart style layout).
    \item \textbf{Enhanced Tutorials and Documentation:} Improve documentation with video tutorials, sample scenes, and detailed integration guides to reduce barriers to entry and facilitate user learning.
\end{itemize}